\documentclass[12pt]{article}

\usepackage[utf8]{inputenc}               
\usepackage[T1]{fontenc}                  
\usepackage[font=small]{caption}          
\usepackage{amsmath, amsthm}              
\usepackage{amsfonts}                     
\usepackage{mathtools}                    
\usepackage{bm, bbm}                      
\usepackage{enumitem}                     
\usepackage{url}                          
\usepackage{float}                        
\usepackage{placeins}                     
\usepackage{xcolor}                       

\usepackage[authoryear,round,comma]{natbib}  

\usepackage{hyperref}                     
\hypersetup{colorlinks=true,citecolor=black,linkcolor=black,urlcolor=black}

\usepackage{authblk}                      

\usepackage{graphicx}                     
\graphicspath{{./fig/}}

\usepackage{geometry}                     
\usepackage{appendix}                     

\newcommand{\beginappendix}{              
    \renewcommand{\thesection}{\Alph{section}}
    \renewcommand{\thesubsection}{\Alph{section}.\arabic{subsection}}
    \setcounter{table}{0}
    \renewcommand{\thetable}{S\arabic{table}}
    \setcounter{figure}{0}
    \renewcommand{\thefigure}{S\arabic{figure}}
}

\usepackage{listings}                     
\definecolor{DarkGreen}{RGB}{1,50,32}
\newtheorem{lemma}{Lemma} 

\newtheorem{proposition}{Proposition}

\theoremstyle{definition}

\newtheorem{remark}{Remark}

\def\E{\mathbb{E}}
\def\Var{\mathrm{Var}}
\def\Cov{\mathrm{Cov}}

\def\bmu{\bm\mu}
\def\hmu{\hat\mu}
\def\hbmu{\hat\bmu}

\def\bbeta{\bm\beta}
\def\hbbeta{\hat\bbeta}

\def\btheta{\bm\theta}
\def\hbtheta{\hat\btheta}

\def\hSigma{\hat\Sigma}
\def\tSigma{\tilde\Sigma}
\def\hsigma{\hat\sigma}

\def\uSigma{\breve\Sigma}

\def\hpsi{\hat\psi}
\def\tpsi{\tilde\psi}
\def\bpsi{\bm\psi}
\def\hbpsi{\hat{\bm\psi}}
\def\tbpsi{\tilde{\bm\psi}}

\def\bg{\mathbf{m}}
\def\hG{\hat{G}}

\def\hQ{\hat{Q}}

\def\hpi{\hat\pi}

\def\hB{\hat{B}}
\def\hM{\hat{M}}
\def\bzero{\mathbf{0}}

\def\hat{\widehat}
\def\tilde{\widetilde}
\def\pr{\mathrm{pr}}
\def\plim{\mathrm{plim}}
\def\aipw{\mathrm{aipw}}
\def\score{\mathrm{score}}

\newcommand{\ES}{\sum_{i=1}^{n}}
\newcommand{\EA}{\frac{1}{n}\sum_{i=1}^{n}}
\newcommand{\deriv}[2]{{\partial{#1}}/{\partial{#2}}}

\begin{document}

\title{\bf A Robust Score Test in G-computation for Covariate Adjustment in Randomized Clinical Trials Leveraging Different Variance Estimators via Influence Functions}

\author[1]{Xin Zhang\thanks{Corresponding author. Email: xin.zhang6@pfizer.com. The last three authors are alphabetically ordered.}}
\author[1,2]{Haitao Chu\thanks{Email: chux0051@umn.edu}}
\author[3]{Lin Liu\thanks{Email: linliu@sjtu.edu.cn}}
\author[1]{Satrajit Roychoudhury\thanks{Email: satrajit.roychoudhury@pfizer.com}}

\affil[1]{Data Sciences and Analytics, Pfizer Inc}
\affil[2]{Division of Biostatistics and Health Data Science, University of Minnesota}
\affil[3]{Institute of Natural Sciences, MOE-LSC, School of Mathematical Sciences, CMA-Shanghai, and SJTU-Yale Joint Center for Biostatistics and Data Science, Shanghai Jiao Tong University}

\date{}

\maketitle

\begin{abstract}
G-computation has become a widely used robust method for estimating unconditional (marginal) treatment effects with covariate adjustment in the analysis of randomized clinical trials.  Statistical inference in this context typically relies on the Wald test or Wald interval, which can be easily implemented using a consistent variance estimator.  However, existing literature suggests that when sample sizes are small or when parameters of interest are near boundary values, Wald-based methods may be less reliable due to type I error rate inflation and insufficient interval coverage.  In this article, we propose a robust score test for g-computation estimators in the context of two-sample treatment comparisons.  The proposed test is asymptotically valid under simple and stratified (biased-coin) randomization schemes, even when regression models are misspecified.  These test statistics can be conveniently computed using existing variance estimators, and the corresponding confidence intervals have closed-form expressions, making them convenient to implement.  Through extensive simulations, we demonstrate the superior finite-sample performance of the proposed method.  Finally, we apply the proposed method to reanalyze a completed randomized clinical trial.  The new analysis using our proposed score test achieves statistical significance, whilst reducing the issue of type I error inflation.
\end{abstract}


\clearpage

\section{Introduction}

In randomized clinical trials, randomization itself ensures unbiased estimation of the unconditional (marginal) treatment effects and allows for valid testing of the null hypothesis of no treatment effect. Nowadays, important patient characteristics are routinely collected at the time of trial enrollment.  Adjusting for these baseline prognostic covariates potentially helps to reduce the conditional bias caused by imbalances in covariates, and may improve the efficiency of statistical analyses. In fact,  recent guidelines from regulatory agencies recommend the use of covariate adjustment \citep{ema2015,fda2023}.

For discrete outcomes, covariate adjustment is typically performed by fitting a nonlinear regression model, such as a generalized linear model (GLM).  An unconditional treatment effect can be estimated by fitting a regression model that includes only an intercept and the treatment indicator variables. However, when baseline covariates are included, the interpretation may shift from an unconditional to a conditional treatment effect \citep{gail1984biased}, even if randomization is properly applied and the model is correctly specified.  

Recently, g-computation \citep{robins1986new} has gained popularity as a method for estimating unconditional treatment effects in randomized clinical trials while adjusting for baseline covariates \citep{ge2011covariate,benkeser2020improving,van2024covariate}.  This approach is also known as regression (model-based or smoothed) standardization \citep{rosenbaum1987model,greenland1991estimating,vansteelandt2011invited}.  It involves fitting an outcome regression model and then averaging the difference of the fitted values for all subjects as if they were assigned to both treatment arms.  For a GLM with the canonical link, randomization alone allows for consistent estimation of unconditional treatment effects, even when the regression model is misspecified \citep{freedman2008randomization,bartlett2018covaraite,guo2023the}. 

Various methods for variance estimation have been proposed for g-computation estimators. One approach involves embedding the g-computation estimator within a system of (stacked) estimating equations, which includes the score equation of a postulated GLM \citep{lunceford2004stratification,yuan2012variable}. This approach views the g-computation estimator as a partial $M$-estimator \citep{stefanski2002the}, and the resulting variance estimator corresponds to one diagonal block of the sandwich variance(-covariance) matrix, accounting for all sources of variation.  \citet{wang2023model} further develop such variance estimators for stratified and biased-coin randomization schemes.

An alternative approach views the g-computation estimator as a semiparametric estimation problem \citep{rosenblum2010simple,ye2023robust}, leveraging efficient influence functions \citep{hampel1974the} to construct estimators. This perspective relies on the equivalence between g-computation estimators using canonical GLMs and certain semiparametric estimators under randomization, which are solutions to estimating equations derived from influence functions. Examples of such estimators include augmented inverse probability weighting (AIPW) \citep{robins1994estimation,tsiatis2008covariate} and targeted maximum likelihood estimation (TMLE) \citep{van2006targeted,moore2009covariate}; see footnote 45 of \citet{richardson2014causal} for some further remarks. Consequently, the resulting variance estimator is simply a plug-in estimator of the variance corresponding to the efficient influence functions, divided by the sample size.

In the absence of model misspecification, both estimators target the same asymptotic variance under simple randomization \citep{tsiatis2006the}. However, it is not immediately clear whether this equivalence holds when the parametric models are misspecified and/or when covariate-adaptive randomization is used. It is important to note that, the latter estimator is generally biased in the nonrandomized settings, when the working model for the nuisance parameter is misspecified \citep{vermeulen2015bias}.

When using g-computation, statistical inference is typically based on the Wald test or Wald confidence interval, along with one of the aforementioned variance estimators. As highlighted in the literature \citep[e.g.,][]{agresti2011score}, Wald tests can perform poorly when sample sizes are small or when parameters of interests are near boundary values, often resulting in inflated type I error rates. Even in trials with large sample sizes, this issue can limit the number of baseline covariates that can be included for adjustment.  Specifically, with a large number of covariates in the model, the ``effective'' sample size becomes smaller, which can lead to substantial estimation bias. 

Two main contributions of this article include: (1) a robust score test for g-computation estimators of unconditional treatment effects in two-sample treatment comparisons, and (2) a novel justification of the variance estimation methods relying on influence functions \citep{rosenblum2010simple,ye2023robust}.  Both are obtained under misspecified parametric models and for simple and stratified (biased-coin) randomization schemes.  The rest of the article is organized as follows.  Section~\ref{sec:pre} introduces g-computation estimators using stacked estimating equations. Section~\ref{sec:theory} presents the theoretical results on variance estimation. In Section~\ref{sec:method}, we develop the test statistics using existing variance estimators for the proposed score test and construct the corresponding confidence intervals.  Performance of the proposed method is evaluated using extensive simulation in Section~\ref{sec:sim}.  Finally, in Section~\ref{sec:trial}, we apply the proposed method to a completed phase 3 randomized trial.  Section~\ref{si:code} in the Supporting Information includes the R code for replication purposes.

\section{Preliminaries} \label{sec:pre}

We consider a randomized clinical trial with $n$ subjects and two arms.  Let Arm $1$ be the reference arm and Arm $2$ be the treatment arm under evaluation. The true randomization proportion for two arms are denoted by $\pi_1 \in (0, 1)$ and $\pi_2 \coloneqq 1 - \pi_1$, respectively. Let $Y_i$ represent the outcome of interest observed for subject $i$ under randomly assigned arm, denoted by $A_i \in \{ 1, 2 \}$. For example, $A_i = 1$ indicates that subject $i$ is assigned to Arm $1$.  Let $W_i$ denote the vector of baseline covariates collected for Subject $i$, which includes important baseline prognostic factors and patient characteristics. Some of these covariates may be used in the randomization process to ensure balance between the arms, depending on the study design.  Finally, we define $D_i \coloneqq ( Y_i, A_i, W_i^\top )^\top$ as the observed data vector for Subject $i$. 

Let $Y_i ( a )$ represent the potential outcome for Subject $i$ under Arm $a$, and let $\mu_a \coloneqq \E [ Y_i ( a ) ]$ be the mean potential outcome for this arm.  The goal is to evaluate the unconditional treatment effect between two arms, denoted by $f ( \bmu )$ where $\bmu \coloneqq ( \mu_1, \mu_2 )^\top$ is the vector of means for the two arms. The treatment effect may be represented by various metrics, such as the difference ($\mu_2 - \mu_1$) or the ratio ($\mu_2 / \mu_1$).  

We assume that the triplet $( Y_i ( 1 ), Y_i ( 2 ), W_i^\top ) ^ \top$ is independent, identically distributed (i.i.d.) and drawn from some superpopulation distribution $\mathbb{P} ( \cdot )$. In the case of simple randomization, this assumption implies that the data vector $D_1, \ldots, D_n$ are also i.i.d. samples. However, the i.i.d assumption does not hold for stratified/biased-coin randomization \citep{kahan2012improper}.  Let $S_i$ denote the stratum assigned to Subject $i$. For stratified/biased-coin randomization, we assume that: (1) conditional independence, $A_i \perp W_i \vert S_i$, meaning the treatment assignment $A_i$ is independent of the baseline covariates $W_i$ given the stratum $S_i$, and (2) the probability of treatment assignment within each stratum is given by $\pr ( A_i = a \vert S_i ) \coloneqq \pi_a$. 

\subsection{G-computation in randomized clinical trials} \label{sec:pre-gc}

The g-computation estimator is constructed by positing a working model for $\E [Y_i \vert A_i = a, W_i]$, where the goal is to model the mean outcome under Arm $a$ given the baseline covariates $W_i$.  A commonly used form for this working model is a GLM with the canonical link function.  Let $m ( \cdot )$ be the inverse of the canonical link function and $I ( \cdot )$ represents the indicator function.  The covariates to be included in $m ( \cdot )$ are  
$X_i \coloneqq ( I ( A_i = 1 ), I ( A_i = 2 ), Z_i^\top )^\top$, which are obtained from $A_i$ and $W_i$.  The first two elements in $X_i$ are dummy variables encoding the treatment arm assignments.  The rest of $X_i$ (i.e., $Z_i$) are obtained from (transformations of) $W_i$, and potentially $A_i$ if including interaction terms for treatment effect heterogeneity.  Then, a working (outcome-regression) model is written as $m ( \bbeta^\top X_i )$.  Both homogeneous and heterogeneous working models can be written in this general form, as shown in Section~\ref{si:wm} of the Supporting Information. This flexibility allows the working model to account for various types of treatment effect variation across subjects. 

Let $X_{i ( a )}$ denote the vector after setting $A_i \equiv a$ in $X_i$.  Formally, the g-computation estimator for the arm-specific mean $\mu_a$ is given by $\hmu_a = \EA m ( \hbbeta^\top X_{i ( a )} )$, where $\hbbeta$ is the estimator of $\bbeta$ from the working model.  The estimated unconditional treatment effect between two arms is then $f(\hbmu)$, where $\hbmu \coloneqq ( \hmu_1, \hmu_2 )^\top$.

Under randomization, $\hbmu$ is consistent for $\bmu$ even when the working model, $m ( \bbeta^\top X_i )$, is completely misspecified \citep{freedman2008randomization,bartlett2018covaraite}. It relies on the ``prediction unbiasedness'' property of GLMs with canonical links \citep{guo2023the}. However, this property does not generally hold for other regression models.  Therefore, in this article, we restrict our focus to canonical GLMs, where this consistency property is well-established and provides the theoretical foundation for our approach.

\subsection{G-computation and estimating equations}  \label{sec:pre-mest}

One variance estimation approach for g-computation is based on $M$-estimation with nuisance parameters \citep{lunceford2004stratification,yuan2012variable}.  This approach utilizes the stacked estimating equations for both $\hbmu$ and $\hbbeta$, where the latter is a nuisance parameter.  More specifically, $\hmu_a$ can be expressed as the solution to the following equation:  
\begin{equation}
\ES \{ m ( \hbbeta^\top X_{i ( a )} ) - \hmu_a \} \equiv 0. \label{eq:ee-mu}  
\end{equation}
The estimator of $\bbeta$ can be constructed using maximum likelihood estimation (MLE), and this estimator ($\hbbeta$) satisfies the standard score equation:
\begin{equation}
    \ES X_i \{ Y_i - m ( \hbbeta^\top X_i ) \} = 0. \label{eq:ee-beta}
\end{equation}
Stacking \eqref{eq:ee-beta} with \eqref{eq:ee-mu} for $a = 1 , 2$ leads to a system of equations that simultaneously estimates both the arm-specific means and the nuisance parameter.  Thus, $\hbmu$ is the so-called partial $M$-estimator \citep{stefanski2002the}, with the variance estimation relying on the sandwich variance(-covariance) estimator. This approach accounts for both model misspecification and the variability in the nuisance parameter estimation.

Specifically, let $\hbtheta \coloneqq ( \hbmu^\top, \hbbeta^\top )^\top$, which is an $M$-estimator as it is the solution to the stacked estimating equations combining \eqref{eq:ee-mu} and \eqref{eq:ee-beta}. The sandwich variance estimator for $\hbtheta$ under simple randomization is given by $\hSigma_{\btheta} = \hB_{\btheta}^{-1} \hM_{\btheta}  \hB_{\btheta}^{-\top} / n$ \citep{stefanski2002the}, where $\hB_{\btheta}$ and $\hM_{\btheta}$ are the empirical estimators for the bread ($B_{\btheta}$) and the meat ($M_{\btheta}$) matrix, respectively. The variance estimator for $\hbmu$, denoted as $\hSigma_{\bmu}$, is then the corresponding diagonal block in $\hSigma_{\btheta}$.  Furthermore, the variance estimator for $f ( \hbmu )$ under simple randomization can be obtained using the delta method, $\{ \deriv{f ( \hbmu )}{\bmu} \} \hSigma_{\bmu} \{ \deriv{f ( \hbmu )}{\bmu} \}^\top$, where $\deriv{f ( \hbmu )}{\bmu} = ( \deriv{f ( \hbmu )}{\mu_1}, \deriv{f ( \hbmu )}{\mu_2})$.

\begin{remark}
\label{rem:wang}
\citet{wang2023model} further develop the variance estimator of $f(\hbmu)$ accounted for stratified and biased-coin randomization schemes using the stacked estimating equations for $f(\hbmu)$ (directly, instead of those for $\hbmu$) and $\hbbeta$.  This differs from the approach described in this section, which first obtains the variance estimator for $\hbmu$ from the stacked estimating equations for $\hbmu$ and $\hbbeta$, and then obtains the variance estimator of $f(\hbmu)$ using the delta method.  Notwithstanding, two approaches are equivalent, as the delta method can also be implemented via $M$-estimation \citep{stefanski2002the}.  We starts with $\hbmu$ as it leads to a coherent procedure to provide analysis results for both $\hbmu$ and $f(\hbmu)$.
\end{remark}

\subsection{G-computation and AIPW} \label{sec:pre-aipw}

It is well-known that, under randomization with canonical GLMs, g-computation estimators solve AIPW estimating equitation \citep{robins2007comment,moore2009covariate}.  This gives rise to an alternative variance estimator for $\hbmu$ under simple randomization.  Specifically, this variance estimator is  $\tSigma_{\bmu} \coloneqq \hat{\Var} [ \tbpsi_{\bmu} ( D_i ) ] / n $, where $\hat{\Var}$ is the sample variance, $\tbpsi_{\bmu} ( D_i ) = \left( \tpsi_1 ( D_i ), \tpsi_2 ( D_i ) \right)^\top$, and
\begin{equation}
    \tpsi_a ( D_i ) = \frac{ I( A_i = a ) }{ \hpi_a } \left\{ Y_i - m ( \hbbeta^\top X_i ) \right\} + m ( \hbbeta^\top X_{i ( a )} ) - \hmu_a, \label{eq:eif-mu-est}
\end{equation}
with $\hpi_a$ being the empirical proportion of Arm $a$.

In the literature, $\tbpsi_{\bmu} ( D_i )$ is a plugin estimator of the AIPW-style influence function for $\hbmu$, and we denote such an influence function by $\bpsi_{\bmu}^\aipw ( D_i ) = ( \psi_1^\aipw ( D_i ), \psi_2^\aipw ( D_i ) )^\top$, where  $\psi_a^\aipw ( D_i ) = \frac{I ( A_i = a )}{\pi_a}\{ Y_i - m ( \bbeta_0^\top X_i ) \} + m ( \bbeta_0^\top X_{ i ( a ) } ) - \mu_a$ \citep{robins1994estimation}.  Provided that $m ( \bbeta^\top X_{i(a)} ) \equiv \E [ Y_i \vert A_i=a, W_i ]$ (i.e., the working model is correctly specified), we have that $\sqrt{n} (\hbmu - \bmu) = \frac{1}{\sqrt{n}} \ES \bpsi_{\bmu}^\aipw ( D_i ) + o_{P} (1)$, where $\E [ \bpsi_{\bmu}^\aipw ( D_i ) ] \equiv \bzero$ and $\E [\bpsi_{\bmu}^\aipw ( D_i )\bpsi_{\bmu}^\aipw ( D_i )^\top ]$ is positive semidefinite, and $\bpsi_{\bmu}^\aipw ( D_i )$ is then the so called efficient influence function for $\hbmu$.  In nonrandomized settings with misspecified parametric models, \citet{vermeulen2015bias} show that $\bpsi_{\bmu}^\aipw ( D_i )$ is generally biased, leading to underestimation by the usual variance estimator $\tSigma_{\bmu}$.  Under randomization, the bias terms vanish, implying that it is a valid influence function for $\hbmu$, and then the variance estimator $\tSigma_{\bmu}$ is consistent.

\begin{remark}
\label{rem:ye}
\citet{ye2023robust} propose a modified variance estimator for $\hbmu$ by expanding $\Var [ \bpsi_{\bmu}^\aipw ( D_i ) ]$.  We denote their estimator as $\uSigma_{\bmu}$ and provide its formula in Section~\ref{si:var-ye} of the Supporting Information.  The expansion of $\Var [ \bpsi_{\bmu}^\aipw ( D_i ) ]$ involves $\Var [ m ( \bbeta_0^\top X_{i(a)} ) \vert A_i = a ]$ ($a=1,2$) and $\Cov [ m ( \bbeta_0^\top X_{i(a)} ), m ( \bbeta_0^\top X_{i(b)} ) \vert A_i = a ]$ ($a \neq b$), which are replaced by the corresponding unconditional ones in their estimator, by leveraging $A_{i} \perp W_{i}$ under simple randomization.  This enables the use of data from two arms to compute certain conditional variance/covariance matrices.
\end{remark}

\begin{remark}
\label{rem:avar}
Provided that $m ( \bbeta^\top X_{i(a)} ) \equiv \E [ Y_i \vert A_i = a, W_i ]$ (i.e., the working model is correctly specified), all three variance estimators of $\hbmu$ introduced in Section~\ref{sec:pre} are targeting the same asymptotic variance under simple randomization \citep{tsiatis2006the}.  It is not immediately clear whether such an equivalence still holds under misspecified parametric models or for stratified/biased-coin randomization, which we will investigate in Section~\ref{sec:theory}.
\end{remark}

\section{Theory: Unifying Variance Estimators} \label{sec:theory}

Let $\Sigma_{\btheta}$ be the asymptotic variance of $\hbtheta$ under simple randomization, associated with the sandwich variance estimator $\hSigma_{\btheta}$.  The asymptotic variance of $\hbmu$, denoted by $\Sigma_{\bmu}$, is then the diagonal block extracted from $\Sigma_{\btheta}$ corresponding to $\bmu$, with $\hSigma_{\bmu}$ (introduced in Section~\ref{sec:pre-mest}) being its estimator.  Let $\bbeta_0 \coloneqq \plim ( \hbbeta )$ as $n \to \infty$ (throughout this article $\plim$ is reserved for probability limit) and $m^\prime ( \cdot )$ being the first derivative of $m ( \cdot )$.  The influence function of $\hbbeta$ is $\bpsi_{\bbeta} ( D_i ) \coloneqq  B_{\bbeta}^{-1} X_i \{ Y_i - m ( \bbeta_0^\top X_i ) \}$, where $B_{\bbeta} \coloneqq \E [ m^\prime ( \bbeta^\top_0 X_i ) X_i X_i^\top ]$ \citep{stefanski2002the}.  It satisfies that $\sqrt{n} (\hbbeta - \bbeta_0) = \frac{1}{\sqrt{n}} \ES \bpsi_{\bbeta} ( D_i ) + o_{P} (1)$, and $\Var [ \bpsi_{\bbeta} ( D_i ) ] = B_{\bbeta}^{-1} M_{\bbeta} \{ B_{\bbeta}^{-1} \}^\top$, where $M_{\bbeta} \coloneqq \E [ \{ Y_i - m ( \bbeta^\top_0 X_i ) \}^2 X_i X_i^\top ]$. 

First of all, we obtain the influence function for $\hbmu$, denoted by $\bpsi_{\bmu}^\score ( D_i )$, from $\Sigma_{\btheta}$ under simple randomization (see Section~\ref{si:eif-mu-mest} of the Supporting Information), which satisfies that $\sqrt{n} (\hbmu - \bmu) = \frac{1}{\sqrt{n}} \ES \bpsi_{\bmu}^\score ( D_i ) + o_{P} (1)$ and $\Var [ \bpsi_{\bmu}^{\score} ( D_i ) ] / n = \Sigma_{\bmu}$.  Specifically, $\bpsi_{\bmu}^\score ( D_i ) = ( \psi_1^\score ( D_i ), \psi_2^\score ( D_i ) )^\top$, where
\begin{equation}
    \psi_a^\score ( D_i ) = \E \left[ m^\prime ( \bbeta_0^\top X_{i ( a )} ) X_{i ( a )} \right]^\top \bpsi_{\bbeta} ( D_i ) + m ( \bbeta_0^\top X_{i ( a )} ) - \mu_a, \label{eq:if-mu}
\end{equation}
is the influence function of $\hmu_a$.   It is the ``score-based'' influence function \citep{fisher2021visually}, which involves the score equation of the working model through $\bpsi_{\bbeta} ( D_i )$.   Finally, Theorem~1 in \citet{wang2023model} concludes that $\bpsi_{\bmu}^\score ( D_i )$ is also the influence function of $\hbmu$ under stratified and biased-coin randomization. 

\citet{bartlett2018covaraite} obtains the asymptotic variance for $\hmu_a$ by considering it as a two-step estimator \citep{newey1994large}, which coincides with $\Var [ \psi_a^\score ( D_i ) ] / n$.  The author does not study the covariance between $\hmu_1$ and $\hmu_2$, which is necessary to obtained the asymptotic variance for $f ( \hbmu )$.  We directly obtain the influence function for $\hbmu$ (of both arms) leading to $\Sigma_{\bmu}$.  \citet{wang2023model} studies the asymptotic variance of $f ( \hbmu )$ directly.  It also involves the influence function, for which however, they do not provide an analytical formula.

\subsection{Equivalence under model misspecification} \label{sec:avar-equiv}

By far, we have introduced two types of influence functions for g-computation estimators, one is $\bpsi^\score ( D_i )$ in \eqref{eq:if-mu} and the other is the AIPW-style influence function $\bpsi^\aipw ( D_i )$ introduced in Section~\ref{sec:pre-aipw}.  They have distinct formulas as the latter one only involves the working model itself, not its score function.  Under simple randomization and assuming that the working model is correctly specified, \citet{bartlett2018covaraite} has shown that the two influence functions are equivalent.  In the following, we show that such equivalence remains true under model misspecification for simple and stratified (biased-coin) randomization.  

We first have the following lemma, which is a direct consequence of the types of experimental designs considered in this work (see Section~\ref{si:strata} of the Supporting Information for the proof). 
\begin{lemma} \label{lem:strata}
    Under stratified/biased-coin randomization, $\pr ( S_i = s \vert A_i = a ) = \pr ( S_i = s )$ for $a \in \{1,2\}$ and every $s$.
\end{lemma}
\noindent Armed with Lemma~\ref{lem:strata}, we have the following result that formally equates $\psi_a^\score ( D_i )$ with $\psi_a^\aipw ( D_i )$ under misspecified parametric working models (see Section~\ref{si:eif-mu-semiparam} of the Supporting Information for the proof). 
\begin{proposition}\label{thm:eif-mu}
    Under simple and stratified (biased-coin) randomization, for $a \in \{ 1, 2 \}$,
    \begin{equation}
        \psi_a^\score ( D_i ) \equiv \psi_a^\aipw ( D_i ) \coloneqq \frac{I ( A_i = a )}{\pi_a}\{ Y_i - m ( \bbeta_0^\top X_i ) \} + m ( \bbeta_0^\top X_{ i ( a ) } ) - \mu_a. \label{eq:eif-mu}
    \end{equation}
\end{proposition}
\noindent The equivalence revealed in Proposition~\ref{thm:eif-mu} indicates that under simple randomization 
\[
    \Var [ \bpsi_{\bmu}^{\aipw} ( D_i ) ] / n \equiv \Var [ \bpsi_{\bmu}^{\score} ( D_i ) ] / n = \Sigma_{\bmu},
\] 
that is, the asymptotic variance for $\hbmu$ obtained from $\Sigma_{\btheta}$ based on $M$-estimation is equal to the one based on the AIPW-stype influence function, even when the working model is misspecified.  As the asymptotic variance under stratified/biased-coin randomization is also characterized by the influence function \citep{wang2023model}, such equivalence continues to hold for these two randomization schemes.

Finally, we show in the following proposition that the asymptotic variance of $\hbmu$ developed by \citet{ye2023robust} (see Remark~\ref{rem:ye}) is identical to $\Sigma_{\bmu}$ under simple and stratified (biased-coin) randomization with model misspecification (see Section~\ref{si:avar-ye} of the Supporting Information for the proof).  
\begin{proposition}\label{thm:var-dec}
Let $\Sigma_{\bmu} ( a, b )$ be the $( a, b )$-cell of $\Sigma_{\bmu}$.  For simple and stratified (biased-coin) randomization, given any $a, b \in \{1, 2\}$, 
\[
    \Sigma_{\bmu} ( a, b ) = \ 
    \begin{cases}
        \begin{aligned} \frac{1}{n \pi_a} \Var \left[ Y_i ( a ) - m ( \bbeta_0^\top X_{i ( a )} ) \right] + \frac{2}{n} \Cov \left[ Y_i(a), m ( \bbeta_0^\top X_{i ( a )} ) \right] \\ - \frac{1}{n} \Var\left[ m ( \bbeta_0^\top X_{i ( a )} ) \right] \end{aligned} & a = b,  \\
        \begin{aligned} \frac{1}{n} \Cov \left[ Y_i ( a ), m ( \bbeta_0^\top X_{i ( b )} ) \right] + \frac{1}{n} \Cov \left[ Y_i ( b ), m ( \bbeta_0^\top X_{i ( a )} ) \right] \\ - \frac{1}{n} \Cov \left[ m ( \bbeta_0^\top  X_{i ( a )} ), m ( \bbeta_0^\top X_{i ( b )} ) \right] \end{aligned}  & a \neq b.
    \end{cases}
\]
\end{proposition}

\subsection{Decomposing asymptotic variance} \label{sec:var-dec}

Our next objective is to study the source of variations of $\hbmu$ by decomposing $\Sigma_{\bmu}$ into three components upon expansion of $\bpsi_{\bmu}^\score ( D_i ) \bpsi_{\bmu}^\score ( D_i )^\top$. 

The first component is $G_{\bbeta} \Sigma_{\bbeta} G_{\bbeta}^\top$, where $G_{\bbeta}$ is a $2 \times p$ ($p$ is the dimension of $X_i$) matrix with the $a$th row being $\E [ m^\prime ( \bbeta_0^\top X_{i ( a )} ) X_{i ( a )}^\top ]$.  As $\Sigma_{\bbeta}$ is the asymptotic variance of $\hbbeta$, this component explains the variation of $\hbmu$ due to the estimation of $\bbeta$.  Note that the variance estimator obtained using the delta method proposed by \citet{ge2011covariate} (if using the sandwich variance estimator for $\hbbeta$) is consistent with this component.  This method has been criticized, from the superpopulation perspective, to underestimate the variability without considering the randomness of $W_i$ \citep{bartlett2018covaraite,ye2023robust}. 

The second component is $\E [ \{ \bg ( W_i; \bbeta_0 ) - \bmu \} \{ \bg ( W_i; \bbeta_0 ) - \bmu \}^\top ] / n$ with $\bg ( W_i; \bbeta_0 ) = \left( m ( \bbeta_0^\top X_{i ( 1 )} ), m ( \bbeta_0^\top X_{i ( 2 )} ) \right)^\top$.  As $\E [ \bg ( W_i; \bbeta_0 ) ] = \bmu$, $\E [ \{ \bg ( W_i; \bbeta_0 ) - \bmu \} \{ \bg ( W_i; \bbeta_0 ) - \bmu \}^\top ]$ is the variance of $\bg ( W_i; \bbeta_0 )$.  This component explains the variation of $\hbmu$ due to the variability of the baseline covariates $W_i$. 

The third component is $\E [ G_{\bbeta} \bpsi_{\bbeta} ( D_i ) \{ \bg ( W_i; \bbeta_0 ) - \bmu \}^\top ] / n$ plus its transpose.  \citet{bartlett2018covaraite} shows that this component can be further simplified as 
\[
    \Cov \left\{ \E [ G_{\bbeta} \bpsi_{\bbeta} ( D_i ) \vert A_i, W_i ], \E [ \bg ( W_i; \bbeta_0 ) - \bmu \vert A_i, W_i] \right\}. 
\]
This term reduces to a matrix of zeros when the working model is correctly specified. Thus, the third component explains certain variability due to model misspecification. 

\subsection{Variance estimation} \label{sec:var-est}

Finally, we summarize the three aforementioned variance estimators for $\hbmu$ (see Table~\ref{tab:var}).  Though they are estimators with the same asymptotic variance as shown in Section~\ref{sec:avar-equiv}, their finite-sample performances are different as implied by  Proposition~\ref{thm:eif-mu}~and~\ref{thm:var-dec}. 

The first one in Table~\ref{tab:var} ($\hSigma_{\bmu}$ introduced in Section~\ref{sec:pre-mest}) is simply extracted from $\hSigma_{\btheta}$, the sandwich variance estimator of $\hbtheta$ \citep{lunceford2004stratification,yuan2012variable}.  Let $\hbpsi_{\bmu} ( D_i ) = ( \hpsi_1 ( D_i ), \hpsi_2 ( D_i ) )^\top$ be a plug-in estimator of $\bpsi_{\bmu}^\score ( D_i )$ with 
\begin{equation}
     \hpsi_a ( D_i ) = \left( \EA m^\prime ( \hbbeta^\top X_{i ( a )} ) X_{i ( a )} \right)^\top \hat{B}_{\bbeta}^{-1} X_i \left\{ Y_i - m ( \hbbeta^\top X_i ) \right\} + m ( \hbbeta^\top X_{i ( a )} ) - \hmu_a. \label{eq:if-mu-est} 
\end{equation}
Then, $\hSigma_{\bmu}$ can be conveniently obtained as the sample variance of $\hbpsi_{\bmu} ( D_i )$ divided by $n$.  This computation procedure is much simpler, only requiring the first derivative of $m(\cdot)$ and the rest are available from the output of fitting a GLM using off-the-shelf software packages. 

The second one in Table~\ref{tab:var} ($\tSigma_{\bmu}$ introduced in Section~\ref{sec:pre-aipw}) is the well-known variance estimator using the AIPW-style (or the so called efficient) influence function \citep{moore2009covariate,rosenblum2010simple}.  Several simulation studies \citep{moore2011robust,tackney2023} show that it may perform poorly when the working model is severely misspecified and fits the data poorly.  This estimator is numerically different from $\hSigma_{\bmu}$ depending on the estimation variability of $\hbbeta$.  When the variability of $\hbbeta$ is small (i.e., $\hbbeta \approx \bbeta_0$), $\tSigma_{\bmu} \approx \hSigma_{\bmu}$ as $\hpsi_a ( D_i ) \approx \tpsi_a ( D_i )$ (see Proposition~\ref{thm:eif-mu}).  But $\tSigma_{\bmu}$ is numerically more stable as its does not involve the inversion of $\hat{B}_{\bbeta}$ as $\hSigma_{\bmu}$ does.  On the other hand, the approximated equality between $\tSigma_{\bmu}$ and $\hSigma_{\bmu}$ no longer holds if the variability of $\hbbeta$ is large (thus $\hbbeta$ is mostly very different from $\bbeta_0$).  Under such circumstances, $\tSigma_{\bmu}$ is a poor variance estimator, which tends to underestimate the true variability of $\hbmu$ (as it ignores the estimation variability of $\hbbeta$, while $\hSigma_{\bmu}$ takes this part into account). 

The last one in Table~\ref{tab:var} ($\uSigma_{\bmu}$ introduced in Remark~\ref{rem:ye}) is the variance estimator proposed by \citet{ye2023robust}.  It inherits the features of $\tSigma_{\bmu}$ as both are built upon $\bpsi_{\bmu}^\aipw ( D_i )$. Further, $\uSigma_{\bmu}$ has the additional advantage in using data from two arms to compute certain conditional variance/covariance matrices, which further improves the numerical stability over $\tSigma_{\bmu}$ (see Remark~\ref{rem:ye}). 

We have conducted a preliminary simulation study to evaluate the performance of these three variance estimators.  See Section~\ref{si:sim-var} of the Supporting Information for the detail.  All estimators have almost the same relative efficiency when sample sizes are sufficiently large.  In most scenarios, $\tSigma_{\bmu}$ enjoys the largest degree of efficiency improvement on average.  However, this may be due to the underestimation of the true variability when sample sizes are small.  Besides, when adjusting for weakly associated covariates, the g-computation estimators are likely to harm the efficiency, as shown in Figure~\ref{fig:re} of the Supporting Information. 

\begin{table}[t!]
    \centering
    \begin{tabular}{l|c|c}
    \hline
             Estimators & Formulas & Influence Functions \\ 
             \hline
      $\hSigma_{\bmu}$ (I) & $\hat{\Var}[(\hpsi_1(D_i),\hpsi_2(D_i))^\top]/n$;  & $\bpsi_{\bmu}^\score ( D_i ) = ( \psi_1^\score ( D_i ), \psi_2^\score ( D_i ) )^\top$ \\
      $\tSigma_{\bmu}$ (II) &  $\hat{\Var}[(\tpsi_1(D_i),\tpsi_2(D_i))^\top]/n$ & $\bpsi_{\bmu}^\aipw ( D_i ) = ( \psi_1^\aipw ( D_i ), \psi_2^\aipw ( D_i ) )^\top$ \\
      $\uSigma_{\bmu}$ (III) & Section~\ref{si:var-ye} in the SI & Same as II \\
      \hline
    \end{tabular}
    \caption{Summary of the three variance estimators for $\hbmu$ descried in Section~\ref{sec:var-est} ($\hat{\Var}$: sample variance; $\hpsi_a(D_i)$: \eqref{eq:if-mu-est}; $\tpsi_a(D_i)$: \eqref{eq:eif-mu-est}; SI: Supporting Information.}
    \label{tab:var}
\end{table}

\section{Methodology: Generalized Score Tests} \label{sec:method}

Wald tests and confidence intervals are typically used with g-computation estimators.  They are convenient to implement using any of the variance estimators displayed in Table~\ref{tab:var}.  Take $H_D: \mu_2 - \mu_1 = \delta_0$ as an example.  A variance estimator for $\hmu_2 - \hmu_1$ can be obtained from $\hSigma_{\bmu}$ (using the delta method), written as $\hsigma_D^2 = \hSigma_{\bmu}(2,2) - 2\hSigma_{\bmu}(2,1) + \hSigma_{\bmu}(1,1)$.  Then, the Wald $\chi^2$ statistic is simply $ ( \hmu_2 - \hmu_1 - \delta_0 )^2 / \hsigma_D^2 $ and the corresponding $(1-\alpha)\times100\%$ confidence interval is 
$\mu_2 - \mu_1 \in \left[ \hmu_2 - \hmu_1 - \hsigma_D z_{1-\alpha/2}, \hmu_2 - \hmu_1 + \hsigma_D z_{1-\alpha/2} \right]$,
where $z_{1-\alpha/2}$ is the upper $1-\alpha/2$ quantile of the standard normal distribution. 

In this section, we propose an alternative method for hypothesis testing and interval estimation within the superpopulation framework for g-computation estimators.  The proposed method is built upon the generalized score test \citep{boos1992on} developed for $M$-estimators.  We have developed the generalized score statistics for testing differences in means and ratios of means under simple randomization (see Section~\ref{si:gscore-stats-sr} in the Supporting Information for the detail).  Besides, we show that those statistics are also valid under stratified/biased-coin randomization (see Section~\ref{si:gscore-stats-strat} in the Supporting Information for the proof).  The proposed test statistics and the corresponding confidence intervals can be conveniently computed, as does the Wald-based method,  using existing variance estimators (Table~\ref{tab:var}). Such statistics and confidence intervals for other types of treatment effect measures can be obtained analogously.  

We first present the proposed method for $H_D$, and the one for $H_R: \mu_2 /\mu_1 = \delta_0$ is provided in Section~\ref{si:ratio} in the Supporting Information.  The generalized score statistic for $H_D$ is  
\begin{align}
    \hQ_D = \frac{ ( \hmu_2 - \hmu_1 - \delta_0 )^2 }{ \hsigma_D^2 + ( \hmu_2 - \hmu_1 - \delta_0 )^2 / n }. \label{eq:score-stats-diff} 
\end{align}
The first term in the denominator ($\hsigma_D^2$) is the variance estimator for $\hmu_2 - \hmu_1$, which can be calculated using any of $\hSigma_{\bmu}$, $\tSigma_{\bmu}$ and $\uSigma_{\bmu}$ from Table~\ref{tab:var}, as a direct consequence of Proposition~\ref{thm:eif-mu}~and~\ref{thm:var-dec}.  The proposed $\hQ_D$ is almost the same as the Wald chi-squared statistic, except for the second term in the denominator, $(\hmu_2 - \hmu_1 - \delta_0)^2/n$.  This additional term ensures that the value of $\hQ_D$ is always smaller than the Wald $\chi^2$ statistic.  Besides, it measures the discrepancy between $\hmu_2 - \hmu_1$ and $\delta_0$.  When the null is true and the working model is well fitted, this term is close to zero, and thus $\hQ_D$ becomes close to the corresponding Wald statistic.  On the contrary, when the working model is poorly fitted, $\hmu_2 - \hmu_1$ may move away from $\delta_0$, and thus, the null hypothesis is likely to be falsely rejected if using the Wald test.  The additional term automatically penalizes $\hQ_D$, leading to a smaller test statistic, and thus reduces the chance of erroneous rejection. 

The corresponding $(1-\alpha)\times100\%$ confidence interval, constructed by inverting the proposed score test, is written as
\begin{equation}
    \mu_2-\mu_1 \in \left[ \hmu_2 - \hmu_1 - \hsigma_D \sqrt{\frac{\chi^2_{1-\alpha}}{1-\chi^2_{1-\alpha}/n}}, \hmu_2 - \hmu_1 + \hsigma_D \sqrt{\frac{\chi^2_{1-\alpha}}{1-\chi^2_{1-\alpha}/n}} \right], \label{eq:score-ci-diff}
\end{equation}
where $\chi^2_{1-\alpha}$ is the upper $(1-\alpha)$ quantile of the chi-squared distribution with one degree of freedom, or $\chi^2_{1-\alpha} = z_{1-\alpha/2}^2$.  The above confidence interval has a similar form to the Wald interval, both being symmetric about $\hmu_2 - \hmu_1$.  But it always has a slightly wider range, though the difference will be negligible when the sample size is large. This indicates that the proposed score confidence interval always has a higher coverage probability than the Wald interval.

The proposed test statistic $\hQ_D$ is developed under simple randomization, and it converges in law under $H_D$ to $\chi^{2}_{1}$, the chi-square distribution with one degree of freedom \citep{boos1992on}.  For stratified/biased-coin randomization, such generalized score statistic is also analytically available (see Section~\ref{si:gscore-stats-strat} of the Supporting Information).  Its value is always no smaller than $\hQ_D$.  This implies that the statistical inference based on $\hQ_D$ is conservative, hence asymptotically valid, under these two randomization schemes. 

\section{Simulation studies} \label{sec:sim}

We carry out extensive simulations to investigate the finite-sample operating characteristics of hypothesis testing and interval estimation using g-computation estimators.  The main goals of simulation studies are to: (1) empirically validate the proposed score test and the corresponding confidence interval developed in Section~\ref{sec:method}, and (2) compare the proposed method with the traditional Wald-based method in terms of the finite-sample performance, especially in controlling type I error rates and improving coverage probability of confidence intervals, when adjusting for multiple covariates, with different sample sizes, and model misspecification.

\subsection{Small/moderate sample sizes} \label{sec:sim-small}

We design three hypothetical trials (all with 1:1 randomization) with binary outcomes to evaluate the finite-sample performance of the proposed method. Each trial is powered at $80\%$ with a one-sided significance level of $\alpha=0.025$ to test the null hypothesis $H_0:\mu_2-\mu_1=0$ against their respective alternatives. The three trials are: (1) $N=326$, $\mu_1=30\%$, $H_a: \mu_2-\mu_1=15\%$; (2) $N=326$, $\mu_1=5\%$, $H_a: \mu_2-\mu_1=8.9\%$; and (3) $N=80$, $\mu_1=30\%$, $H_a: \mu_2-\mu_1=30\%$.

The outcomes are drawn from a Bernoulli distribution, with $\pr ( Y_i=1 \vert A_i=a, W_i ) = \text{expit}(\beta_a^A + \beta_1^W W_{i1} + \beta_2^W W_{i2} + \beta_3^W W_{i3})$ where $\text{expit}(x)=1/(1+e^{-x})$. Each of the three baseline covariates ($W_{i1},W_{i2},W_{i3}$) is independently drawn from a standard normal distribution.  Two randomization schemes are applied to generate $A_i\in\{1,2\}$. The first one is designed to approximate simple randomization, while the sizes of the two arms are kept exactly the same, reflecting a completely randomized experiment.  In the second scheme, we apply a stratified permuted block randomization, where subjects are stratified by $S_i \coloneqq I(W_{i3}>0.25)$. The values of the treatment and covariate effects are carefully selected to simulate the three hypothetical trials with moderate overall prognostic effects.  Specifically, for the covariates, we set $\beta_j^W=\sqrt{\log(2)^2/3}$ for $j=1,2,3$. The choice ensures that the cumulative effect of the three covariates on the outcome $Y_i$ is equal to $\log(2)$. The setup is detailed in Section~\ref{si:sim-specs} of the Supporting Information, where additional specifications of other parameters are provided.  In total, we simulate six distinct scenarios, covering a range of situations with varying sample sizes, control arm rates and two commonly used randomization schemes. These scenarios are designed to reflect a broad spectrum of possible real-world settings for randomized clinical trials.

In the simulation study, we compare the performance of the Wald test and the proposed score test, using three variance estimators listed in Table~\ref{tab:var}.  Two g-computation estimators are considered: \textbf{GC(Logit)} and \textbf{GC(Log)}, where the former uses  logistic regression and the latter uses Poisson regression (to explore the impact of model misspecification). Additionally, we include the unadjusted (\textbf{Unadj}) estimators as baseline comparisons, along with the Wald and score tests.  

For simple randomization scenarios, the two g-computation estimators adjust for different sets of covariates: $W_{i1}$, $(W_{i1},W_{i2})$, or $(W_{i1},W_{i2},W_{i3})$. For the stratified randomization scenarios, the adjustments are made for $S_{i}$, $(S_i,W_{i1})$, or $(S_i,W_{i1},W_{i2})$.  

\begin{figure}[t!]
    \centering
    \includegraphics[width=\linewidth]{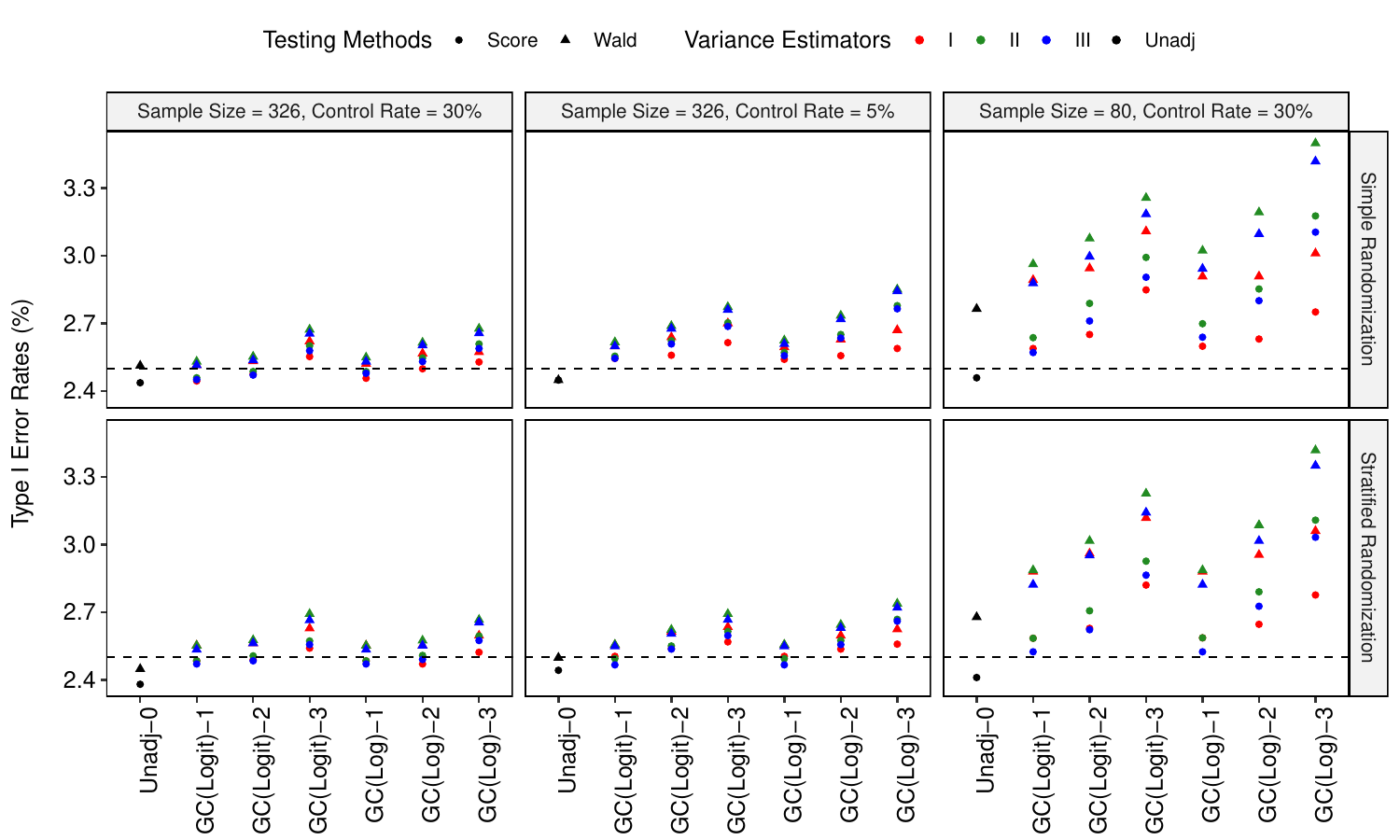}
    \caption{The type I error rate for testing $H_0:\mu_2-\mu_1=0$ using the Wald test and the proposed score test, evaluated for two g-computation estimators, using logistic (\textbf{GC(Logit)}) and Poisson (\textbf{GC(Log)}) regression models, respectively, under three hypothetical trials (by column) and two randomization schemes (by row).  For simple randomization, both g-computation estimators adjust for either $W_{i1}$ (1), $(W_{i1},W_{i2})$ (2), or $(W_{i1},W_{i2},W_{i3})$ (3).  For stratified randomization, they adjust for either $S_{i}$ (1), $(S_i,W_{i1})$ (2), or $(S_i,W_{i1},W_{i2})$ (3).  All three variance estimators, $\hSigma_{\bmu}$ (I), $\tSigma_{\bmu}$ (II), and $\uSigma_{\bmu}$ (III), summarized in Table~\ref{tab:var}, are used in the two testing methods. }
    \label{fig:alpha-mult}
\end{figure}

The simulation results for the type I error rates, based on 50,000 repeated runs for each scenario, are shown in Figure~\ref{fig:alpha-mult}. In all simulations, there were no convergence issue when fitting the working models.   Additionally, the simulation results on interval coverage probabilities under the corresponding alternatives are provided in Section~\ref{si:sim-cp} of the Supporting Information.

Next, we evaluate the power gain of the Wald test and the proposed score test for g-computation estimators with three variance estimators in Table~\ref{tab:var}, compared to the unadjusted tests.  We use \textbf{GC(Logit)} estimator with adjustment for one covariate for simplicity.  The focus is on the hypothetical trial of $N=326$ participants and the control arm outcome of $\mu_1=30\%$, under scenarios where the type I error rates are preserved approximately at the nominal level. 

We investigate three scenarios for covariate distributions. For each of these scenarios, we simulate the baseline covariate $W_i$ as follows: (1) $W_i\in\{0,1\}$, drawn from a Bernoulli distribution with $p_W = 0.1, 0.3, 0.5$; (2) a continuous covariate distribution for $W_i$, drawn from a standard normal distribution. For each scenario, we simulate $(A_i, Y_i)$ as described previously (for simple randomization).  

We test the power of the Wald test and the proposed score test at increasing levels of association between the covariate $W_i$ and the outcome. Specifically, we increase the value of $\exp(\beta^W)$ from 1 to 5, where $\exp(\beta^W)$ (=5) indicates a very strong association, which is uncommon for typical trial populations.  The parameter values for $(\beta_1^A, \beta_2^A)$ are carefully chosen for each value of $\beta^W$ to ensure that the trials are powered at $80\%$ with a one-sided significance level of $\alpha=0.025$ for testing the hypothesis $H_a:\mu_2-\mu_1=15\%$. 

\begin{figure}[t!]
    \centering
    \includegraphics[width=\linewidth]{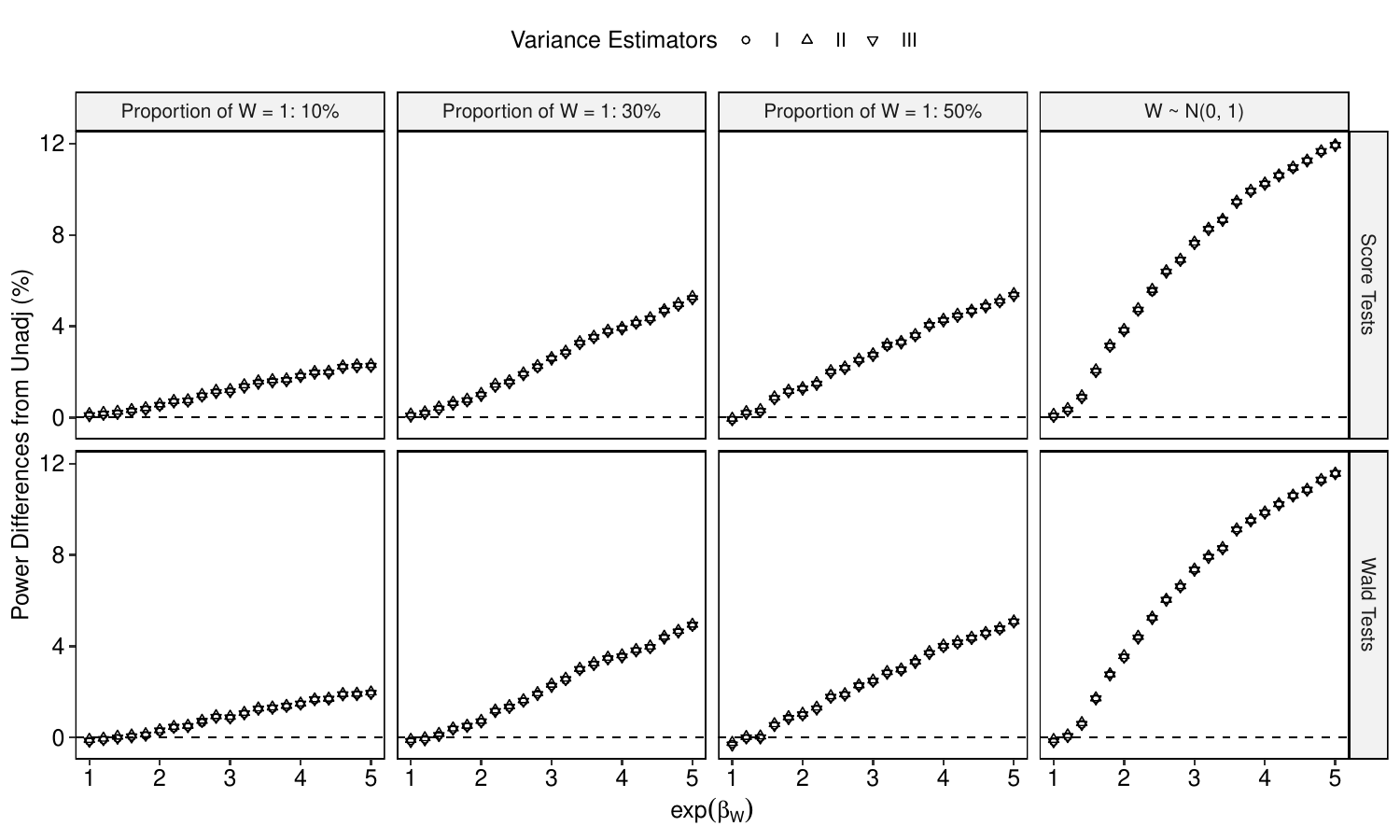}
    \caption{The power difference (minus the power of the unadjusted test) for testing $H_D:\mu_2-\mu_1=0$ using the Wald test and the proposed score test based on \textbf{GC(Logit)} adjusting for $W_i$.  The $(Y, W)$-association is increased by the value of $\exp(\beta_W)$, from 1 to 5 with (a step of 0.2).  Both binary (Column 1-3) and continuous (Column 4) values for $W_i$ are considered.  All three variance estimators, $\hSigma_{\bmu}$ (I), $\tSigma_{\bmu}$ (II), and $\uSigma_{\bmu}$ (III), summarized in Table~\ref{tab:var}, are used in the two testing methods.}
    \label{fig:beta}
\end{figure}

The power gain is defined as the difference between the power of the proposed score test (or the Wald test) and the unadjusted test. This metric quantifies the benefit of adjusting for baseline covariates using the proposed method. For each simulation scenario, we run 50,000 repeated simulations to estimate the power gain under each covariate distribution.  The simulation results for power gain are presented in Figure~\ref{fig:beta}.  There were no convergence issue for fitting working models in all simulations.

\subsection{Large sample sizes} \label{sec:sim-large}

\begin{figure}[t!]
    \centering
    \includegraphics[width=\linewidth]{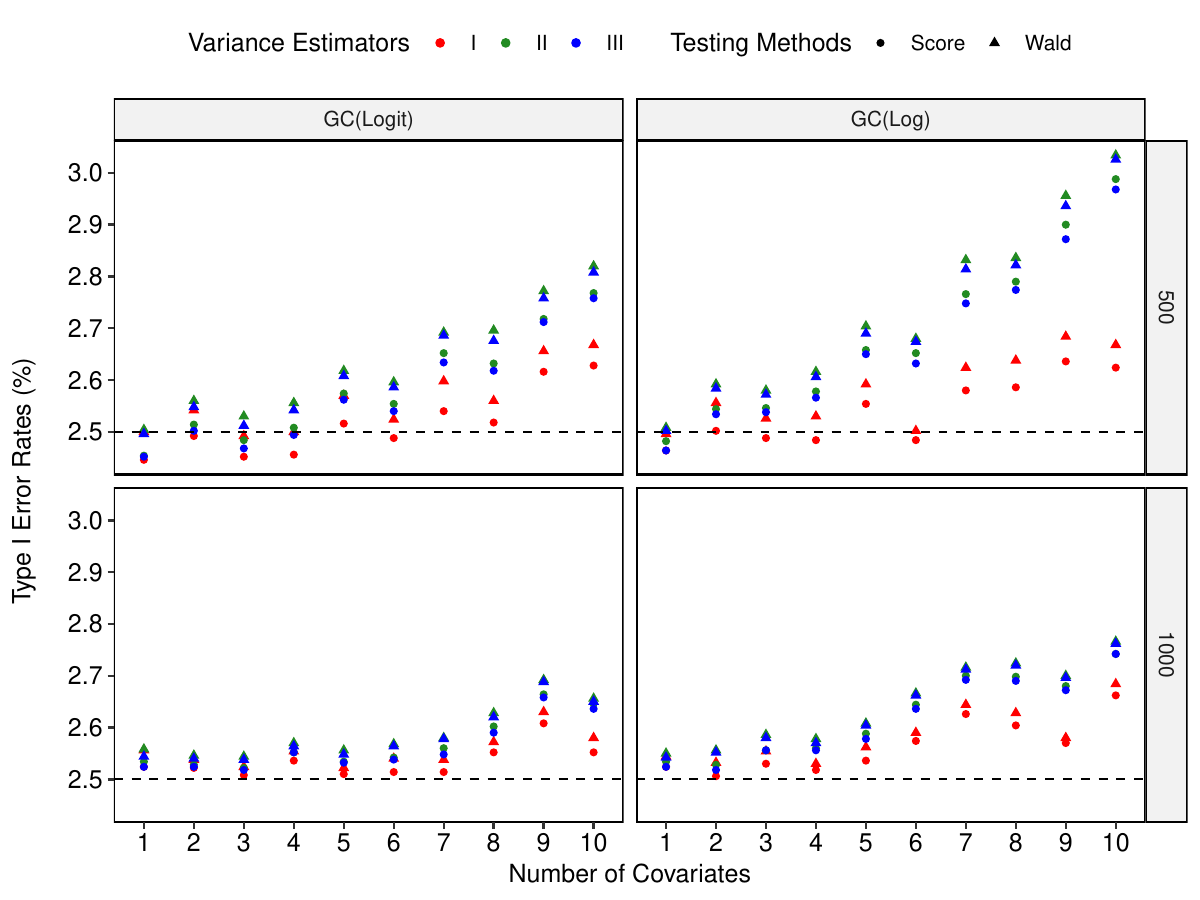}
    \caption{The type I error rate for testing $H_0:\mu_2-\mu_1=0$ using the Wald test and the proposed score test, evaluated for two g-computation estimators (by column), using logistic (GC(Logit)) and Poisson (GC(Log)) regression models, respectively, under two hypothetical trials (by row) of $n=500, 1000$ and $\mu_1=30\%$.  Each g-computation estimator adjusts for $W_{i1}$, $(W_{i1},W_{i2})$, $\dots$, and $(W_{i1},W_{i2},\ldots, W_{i,10})$, respectively.  All three variance estimators, $\hSigma_{\bmu}$ (I), $\tSigma_{\bmu}$ (II), and $\uSigma_{\bmu}$ (III), summarized in Table~\ref{tab:var}, are used in the two testing methods. }
    \label{fig:dim}
\end{figure}

We further evaluate the performance of the Wald test and the proposed score test of g-computation estimators in terms of type I error rates for two hypothetical trials with large sample sizes ($n=500,1000$), adjusting for at most ten baseline covariates.  The simulation procedure is similar to the one described above for simple randomization. Specifically, We set $\beta_j^W=\sqrt{\log(5)^2/10}$ for $j=1,2,\ldots,10$, so that the cumulative effect of those ten covariates on $Y_i$ is equal to $\log(5)$.  The control arm rate is at 30\%.  Figure~\ref{fig:dim} presents type I error rates of the Wald test and the proposed score test for \textbf{GC(Logit)} and \textbf{GC(Log)}.  There are 50,000 repeated runs for each of two scenarios.  There were no convergence issue for fitting working models in all simulations. 

\subsection{Results} \label{sec:sim-res}

Figure~\ref{fig:alpha-mult} shows that the proposed score test outperforms the Wald test in maintaining type I error rates, regardless of whether simple or stratified randomization is used. The score test consistently exhibits lower type I error rates across all variance estimators, than those from the Wald test, particularly when the sample size is small, when the control arm rate is near zero, or when adjusting for multiple covariates.  In the first hypothetical trial, the Wald test, whether using \textbf{GC(Logit)} or \textbf{GC(Log)} with any variance estimators, can adjust for at most two covariates without causing type I error rates to rise to or exceed 0.026.  In contrast, the proposed score test maintains type I error rates below or around 0.026, even when adjusting for up to three covariates.

Additionally, Figure~\ref{fig:alpha-mult} shows that the performance of both Wald and score tests varies depending on the choice of variance estimators.  Those based on $\tSigma_{\bmu}$ experience the largest inflation in type I error rates.  When adjusting for only one covariate, those tests using $\uSigma_{\bmu}$ have similar or slightly smaller type I error rates than those using $\hSigma_{\bmu}$. However, $\hSigma_{\bmu}$ yields the lowest type I error rate when adjusting for more than one covariate. 

The simulation results for interval coverage probabilities of the Wald interval and the proposed score interval are presented in Section~\ref{si:sim-cp} of the Supporting Information. These results align with those shown in Figure~\ref{fig:alpha-mult} for type I error rates, demonstrating similar patterns in performance. The improvement in coverage probabilities is observed consistently in the score confidence intervals.

Figure~\ref{fig:beta} shows that the proposed score test has a similar improvement in the power over the unadjusted score one, comparable to the power gain observed with the Wald test.  For different variance estimators, the power gains are mostly indistinguishable.  Furthermore, we observe that the Wald test for the g-computation estimator has lower power than the unadjusted Wald test when $\exp(\beta_W)$ is small across all scenarios, including when $\exp(\beta_W)=1.4$ in the first scenario. On the other hand, the proposed score test maintains approximately the same power as the unadjusted score test, even when $\exp(\beta_W)=0$.

Additionally, Figure~\ref{fig:beta} highlights that the heterogeneity of $W_i$ influences the magnitude of the power gain through g-computation estimators. In the first three scenarios, when $p_W$ is closer to 0.5 indicating that the heterogeneity of $W_i$ is larger, adjusting for this covariate leads to greater power gain.  This effect is also observed when adjusting continuous covariates: those with higher heterogeneity, after being standardized to $W_i$ (which follows a standard normal distribution), lead to larger values of $\beta_W$, thereby enhancing the power. 

Figure~\ref{fig:dim} shows that in trials with large sample sizes, all testing methods experience significant inflation in type I error rates as the number of adjusted covariates increases, although the magnitude of this inflation varies.  The proposed score test consistently shows lower type I error rates compared to the Wald test. This is consistent with those results presented in Figure~\ref{fig:alpha-mult}.  Those score tests using $\hSigma_{\bmu}$ exhibit the smallest type I error rate inflation.  Interestingly, as the number of adjusted covariates increases, the Wald test using $\hSigma_{\bmu}$ becomes the second best in controlling type I error rates.  

Additionally, for the same testing method based on the same variance estimator, when increasing the number of adjusted covariates, the type I error rates for \textbf{GC(Log)} are larger than those for \textbf{GC(Logit)}. This indicates that model misspecification is likely to contribute to the inflation of type I error rates, which is particularly evident for the hypothetical trial with $n=500$. 

In summary, the simulation results highlight the superiority of the $\hSigma_{\bmu}$-based score test, which allows for the adjustments of more covariates with less inflation in type I error rates in most scenarios.  Adjusting for more baseline covariates can lead to increased power and efficiency gains, but it also carries a higher risk of severe type I error rate inflation and insufficient interval coverage due to potential model over-fitting.

\section{Application} \label{sec:trial}

\begin{table}[t!]
    \centering
    \begin{tabular}{c|ccc}
    \hline
             Estimators & Risk Differences (\%) & 95\% Confidence Intervals (\%) & 1-sided P-values \\\hline
                  Unadj &           9.39 &           (-0.83, 19.62) &          0.0359 \\\hline
                   GC-2 &           9.82 &           (-0.29, 19.94) &          0.0285 \\\hline
                   GC-3 &          10.68 &            (1.92, 19.44) &          0.0085 \\\hline
    \end{tabular}
    \caption{Summary of clinical trial (NIDA-CTN-0003) data analysis results: Unadj = the unadjusted (Wald-based) method; GC-2 = the Wald-based method using g-computation (with $\hSigma_{\bmu}$) adjusting only for the randomization strata; GC-3 = the proposed score-based method using g-computation (with $\hSigma_{\bmu}$) adjusting for the randomization strata and the opioid urine toxicology.}
    \label{tab:app}
\end{table}

The NIDA-CTN-0003 study is a randomized trial completed in 2005 to compare two taper schedules following a period of physiological stabilization on buprenorphine for opioid-dependent individuals \citep{ling2009buprenorphine}.  A total of 516 participants were randomized to either a 28-day (control) or a 7-day taper (treatment) group at the end of the stabilization period.  The randomization was stratified by the maintenance dose (with 3 levels).  The objective of the statistical analysis was to compare the difference in percentages of participants who provided opioid-free urine specimens at the end of the taper across the two taper conditions.  \citet{wang2023model} analyzed a subset of this dataset ($\approx 360$ participants) using the g-computation estimator adjusting for six baseline covariates, which led to significant variance reduction.  

We reanalyze this trial using the g-computation estimator with logistic regression, adjusting for the stratification factor and the baseline opioid urine toxicology result.  In this analysis, we excluded those participants with missing values for either baseline covariates or outcomes. This exclusion resulted in a total of 367 participants, with 167 in the control arm and 200 in the treatment arm.  Let $S_i\in\{1,2,3\}$ be the indicator variable for three randomization strata.  The baseline covariate $W_i$ is the opioid urine toxicology result. The working model is 
$
    \text{expit}\{\beta_1^AI(A_i=1)+\beta_2^AI(A_i=2)+\beta^S_1I(S_i=1)+\beta^S_2I(S_i=2)+\beta^W W_i\}.
$

In this analysis, the working model includes three baseline covariates.  As illustrated in the simulation studies (the fist two columns in Figure~\ref{fig:alpha-mult}), the Wald test for g-computation is able to maintain the type I error rate within a reasonable range when adjusting for at most two covariates, but shows signs of inflation when adjusting for three or more.  On the other hand, the proposed score test, especially the one based on $\hSigma_{\bmu}$, better protects the type I error rate when adjusting for all three baseline covariates.  In other words, using the proposed $\hSigma_{\bmu}$-based score test instead of the Wald test, we can include one more covariate, which allows for a more accurate estimate of the treatment effect and thus potentially increases the likelihood of obtaining a positive result.

Table~\ref{tab:app} displays the estimates of the difference in proportions (7-day taper minus 28-day taper) with the 95\% confidence intervals and one-sided p-values using the following methods: the unadjusted (Wald-based) method (\textbf{Unadj}), the Wald-based method using g-computation adjusting only for the randomization strata (\textbf{GC-2}), and the proposed score-based method using g-computation adjusting for the randomization strata and the opioid urine toxicology result (\textbf{GC-3}).  The results for both \textbf{Unadj} and \textbf{GC-2} are not statistically significant, while, for \textbf{GC-3}, the proposed score test achieves statistical significance. Additionally, the lower bound of its 95\% confidence interval exceed 0 by a non-negligible margin, indicating a strong evidence for a treatment effect.

\section{Discussion} \label{sec:disc}

In this paper, we propose a robust score test for g-computation estimators in randomized clinical trials, specifically for covariate adjustment under simple and stratified (biased-coin) randomization schemes.  The proposed score test is designed to offer better protection against type I error rate inflation compared to the Wald test.  These test statistics are easily computable using those existing variance estimators, which are summarized in Table~\ref{tab:var}.  By inverting the proposed test, we can easily construct the corresponding confidence interval using a closed-form formula.  Through extensive simulation studies, we demonstrate that the proposed method improves over the traditional Wald-based method in terms of finite-sample performance, particularly with lower type I error rates and more precise coverage probabilities, especially when the sample size is small or when the parameter values are extreme (e.g., response rates are close to 0).  Our findings are consistent with previous literature on unadjusted score tests \citep[e.g.,][]{agresti2011score}. We also recommend using the proposed score test in trials with large sample sizes, as it allows for the inclusion of  more baseline covariates with less inflation in type I error rates, potentially leading to greater power gains.

For the three variance estimators, $\hSigma_{\bmu}$, $\tSigma_{\bmu}$ and $\uSigma_{\bmu}$ (Table~\ref{tab:var}), we have \textit{formally} established their asymptotic equivalence (using elementary probability and linear algebra) under simple and stratified (biased-coin) randomization schemes, with misspecified working models (this is shown in Proposition~\ref{thm:eif-mu}~and~\ref{thm:var-dec}). This equivalence further reinforces the superiority of those methods relying on $\hSigma_{\bmu}$.  This is likely due to that $\hSigma_{\bmu}$ directly incorporates the (sandwich) variance of $\hbbeta$, thereby accounting for the additional variability introduce by fitting a misspecified model.  Despite the potential misspecification, $\hSigma_{\bmu}$ retains its robustness and provides reliable variance estimation, thus supporting its optimal performance (for both Wald and score-based methods) in scenarios with small sample sizes, a large number of covariates, or model misspecification, which has been demonstrated in our simulation studies.

An alternative approach to improve finite-sample performance for hypothesis testing and interval estimation is to use a variance estimator with reduced asymptotic bias. For generalized estimating equations (GEE), several bias-corrected estimators have been developed for sandwich variance estimators \citep[e.g.,][]{kauermann2001note,mancl2001covariance}.  These estimators are commonly used to improve the performance of the Wald test and the corresponding confidence interval in finite samples.  They have been applied to g-computation for the Wald-based method to improve finite-sample performance \citep{guo2023the,liu2024covariate}, which can also be applied to the proposed score-based method by simply replacing $\hsigma_D$ in \eqref{eq:score-stats-diff}, \eqref{eq:score-ci-diff} with such estimators.  In Section~\ref{si:sim-bc-var} of the Supporting Information, we provide results from a simulation study that evaluates the performance of both Wald and score tests using several of these alternative variance estimators, including the one with degree-of-freedom adjustment proposed by \citep{tsiatis2008covariate}.  The simulation results show that these potentially less biased estimators can reduce inflation in type I error rates, improving performance in certain settings, for both Wald and score-based methods. 

For stratified and biased-coin randomization schemes, the score-based method proposed in Section~\ref{sec:method} (developed assuming simple randomization) is conservative (i.e., offering strong protection against type I error rates and ensuring reliable interval coverage probabilities).  Alternative generalized score statistics, which are designed to incorporate the correlation induced by stratified/biased-coin randomization, are provided in Section~\ref{si:gscore-stats-strat} of the Supporting Information. Asymptotically, these generalized score statistics are more efficient than the statistics we proposed in Section~\ref{sec:method}.  However, their calculations are significantly more complex, particularly when it comes to constructing the corresponding confidence intervals. Furthermore, they likely experience even more severe inflation in type I error rates in finite samples. Therefore, while they have theoretical advantages in large-sample settings, the proposed score statistics are preferred in practice, especially when dealing with small sample sizes. 

Several directions warrant further investigation. First, multi-arm randomized clinical trials are common with the primary interest often lying in determining pairwise unconditional treatment effects.  While the proposed method is designed for two-arm trials, it is easily extendable to multiple arms under simple randomization.  For stratified/biased-coin randomization, our method is based on approaches developed for two-arm trials \citep{bugni2018inference,wang2023model}. While linear regression adjustment under covariate-adaptive randomization in multi-arm settings has been explored in the literature \citep{bugni2019inference,ye2023toward}, to the best of our knowledge, no work has been done for $M$-estimators in this context. 

Second, multiple testing is commonly conducted in randomized clinical trials, such as comparing multiple treatments to a control, testing multiple outcomes, or assessing multiple subgroups. While the proposed method can control the family-wise error rate using a general multiple testing procedure, more efficient procedures may be developed by combining stacked estimating equations from multiple g-computation estimators \citep{ristl2020simultaneous}.  

Next, missing data is a common challenge in clinical trials and is often addressed by multiple imputation. However, combining imputation methods with g-computation may require special considerations \citep{robins2000inference,seaman2012combining}.  Finally, most existing methods, including ours, relies on the large-sample theory with a fixed number of covariates.  These methods may not perform well when adjusting for more covariates under fixed sample sizes \citep{lei2021regression}. We have also conducted research \citep{zhao2024hoif} in this direction using the framework of higher-order influence functions \citep{robins2016technical}, aiming to address challenges arising from high-dimensional model-free covariate adjustment in finite samples; for recent related works, also see \citet{lu2023debiased} and \citet{chang2024exact}.

\subsection*{Acknowledgments} 

The authors would like to thank Neal Thomas for stimulating discussions, and two anonymous referees and the associate editor for helpful comments.  The dataset used in Section~\ref{sec:trial} is available at \href{https://datashare.nida.nih.gov/study/nida-ctn-0003}{https://datashare.nida.nih.gov/study/nida-ctn-0003}. Lin Liu is also affiliated with the Shanghai Artificial Intelligence Laboratory and his research was supported by NSFC Grant No.12471274 and 12090024 and Shanghai Science and Technology Commission Grant No.21JC1402900 and 2021SHZDZX0102.

\bibliographystyle{apalike}        
\bibliography{main}

\begin{thebibliography}{}

\bibitem[Agresti, 2011]{agresti2011score}
Agresti, A. (2011).
\newblock Score and pseudo-score confidence intervals for categorical data analysis.
\newblock {\em Statistics in Biopharmaceutical Research}, 3(2):163--172.

\bibitem[Bartlett, 2018]{bartlett2018covaraite}
Bartlett, J.~W. (2018).
\newblock Covariate adjustment and estimation of mean response in randomised trials.
\newblock {\em Pharmaceutical Statistics}, 17(5):648--666.

\bibitem[Benkeser et~al., 2020]{benkeser2020improving}
Benkeser, D., D\'{i}az, I., Luedtke, A., Segal, J., Scharfstein, D., and Rosenblum, M. (2020).
\newblock Improving precision and power in randomized trials for {COVID‐19} treatments using covariate adjustment, for binary, ordinal, and time‐to‐event outcomes.
\newblock {\em Biometrics}, 77(4):1467--1481.

\bibitem[Boos, 1992]{boos1992on}
Boos, D.~D. (1992).
\newblock On generalized score tests.
\newblock {\em The American Statistician}, 46(4):327--333.

\bibitem[Bugni et~al., 2018]{bugni2018inference}
Bugni, F.~A., Canay, I.~A., and Shaikh, A.~M. (2018).
\newblock Inference under covariate--adaptive randomization.
\newblock {\em Journal of the American Statistical Association}, 113(524):1784--1796.

\bibitem[Bugni et~al., 2019]{bugni2019inference}
Bugni, F.~A., Canay, I.~A., and Shaikh, A.~M. (2019).
\newblock Inference under covariate‐-adaptive randomization with multiple treatments.
\newblock {\em Quantitative Economics}, 10(4):1747--1785.

\bibitem[Chang et~al., 2024]{chang2024exact}
Chang, H., Middleton, J.~A., and Aronow, P.~M. (2024).
\newblock Exact bias correction for linear adjustment of randomized controlled trials.
\newblock {\em Econometrica}, 92(5):1503--1519.

\bibitem[{European Medicines Agency}, 2015]{ema2015}
{European Medicines Agency} (2015).
\newblock Guidelines on adjustment for baseline covariates in clinical trials.
\newblock \url{https://www.ema.europa.eu/en/documents/scientific-guideline/guideline-adjustment-baseline-covariates-clinical-trials_en.pdf}.
\newblock Accessed Jun 1, 2024.

\bibitem[Fisher and Kennedy, 2021]{fisher2021visually}
Fisher, A. and Kennedy, E.~H. (2021).
\newblock Visually communicating and teaching intuition for influence functions.
\newblock {\em The American Statistician}, 75(2):162--172.

\bibitem[Freedman, 2008]{freedman2008randomization}
Freedman, D.~A. (2008).
\newblock Randomization does not justify logistic regression.
\newblock {\em Statistical Science}, 23(2):237--249.

\bibitem[Gail et~al., 1984]{gail1984biased}
Gail, M.~H., Wieand, S., and Piantadosi, S. (1984).
\newblock Biased estimates of treatment effect in randomized experiments with nonlinear regressions and omitted covariates.
\newblock {\em Biometrika}, 71(3):431--444.

\bibitem[Ge et~al., 2011]{ge2011covariate}
Ge, M., Durham, K.~L., Meyer, D.~R., Xie, W., and Thomas, N. (2011).
\newblock Covariate-adjusted difference in proportions from clinical trials using logistic regression and weighted risk differences.
\newblock {\em Drug information journal : DIJ / Drug Information Association}, 45(4):481--493.

\bibitem[Greenland, 1991]{greenland1991estimating}
Greenland, S. (1991).
\newblock Estimating standardized parameters from generalized linear models.
\newblock {\em Statistics in Medicine}, 10(7):1069--1074.

\bibitem[Guo and Basse, 2023]{guo2023the}
Guo, K. and Basse, G. (2023).
\newblock The generalized {Oaxaca-Blinder} estimator.
\newblock {\em Journal of the American Statistical Association}, 118(541):524--536.

\bibitem[Hampel, 1974]{hampel1974the}
Hampel, F.~R. (1974).
\newblock The influence curve and its role in robust estimation.
\newblock {\em Journal of the American Statistical Association}, 69(346):383--393.

\bibitem[Kahan and Morris, 2012]{kahan2012improper}
Kahan, B.~C. and Morris, T.~P. (2012).
\newblock Improper analysis of trials randomised using stratified blocks or minimisation.
\newblock {\em Statistics in Medicine}, 31(4):328--340.

\bibitem[Kauermann and Carroll, 2001]{kauermann2001note}
Kauermann, G. and Carroll, R.~J. (2001).
\newblock A note on the efficiency of sandwich covariance matrix estimation.
\newblock {\em Journal of the American Statistical Association}, 96(456):1387--1396.

\bibitem[Lei and Ding, 2021]{lei2021regression}
Lei, L. and Ding, P. (2021).
\newblock Regression adjustment in completely randomized experiments with a diverging number of covariates.
\newblock {\em Biometrika}, 108(4):815--828.

\bibitem[Ling et~al., 2009]{ling2009buprenorphine}
Ling, W., Hillhouse, M., Domier, C., Doraimani, G., Hunter, J., Thomas, C., Jenkins, J., Hasson, A., Annon, J., Saxon, A., et~al. (2009).
\newblock Buprenorphine tapering schedule and illicit opioid use.
\newblock {\em Addiction}, 104(2):256--265.

\bibitem[Liu and Xi, 2024]{liu2024covariate}
Liu, J. and Xi, D. (2024).
\newblock Covariate adjustment and estimation of difference in proportions in randomized clinical trials.
\newblock {\em Pharmaceutical Statistics}, 23(6):884--905.

\bibitem[Lu et~al., 2023]{lu2023debiased}
Lu, X., Yang, F., and Wang, Y. (2023).
\newblock Debiased regression adjustment in completely randomized experiments with moderately high-dimensional covariates.
\newblock {\em arXiv preprint arXiv:2309.02073}.

\bibitem[Lunceford and Davidian, 2004]{lunceford2004stratification}
Lunceford, J.~K. and Davidian, M. (2004).
\newblock Stratification and weighting via the propensity score in estimation of causal treatment effects: {A} comparative study.
\newblock {\em Statistics in Medicine}, 23(19):2937--2960.

\bibitem[Mancl and DeRouen, 2001]{mancl2001covariance}
Mancl, L.~A. and DeRouen, T.~A. (2001).
\newblock A covariance estimator for {GEE} with improved small-sample properties.
\newblock {\em Biometrics}, 57(1):126--134.

\bibitem[Moore et~al., 2011]{moore2011robust}
Moore, K.~L., Neugebauer, R., Valappil, T., and {van der Laan}, M.~J. (2011).
\newblock Robust extraction of covariate information to improve estimation efficiency in randomized trials.
\newblock {\em Statistics in Medicine}, 30(19):2389--2408.

\bibitem[Moore and {van der Laan}, 2009]{moore2009covariate}
Moore, K.~L. and {van der Laan}, M.~J. (2009).
\newblock Covariate adjustment in randomized trials with binary outcomes: {Targeted} maximum likelihood estimation.
\newblock {\em Statistics in Medicine}, 28(1):39--64.

\bibitem[Newey and McFadden, 1994]{newey1994large}
Newey, W.~K. and McFadden, D. (1994).
\newblock Large sample estimation and hypothesis testing.
\newblock In {\em Handbook of Econometrics}, volume~4, pages 2111--2245. Elsevier.

\bibitem[Richardson and Rotnitzky, 2014]{richardson2014causal}
Richardson, T.~S. and Rotnitzky, A. (2014).
\newblock Causal etiology of the research of {J}ames {M}. {R}obins.
\newblock {\em Statistical Science}, 29(4):459--484.

\bibitem[Ristl et~al., 2020]{ristl2020simultaneous}
Ristl, R., Hothorn, L., Ritz, C., and Posch, M. (2020).
\newblock Simultaneous inference for multiple marginal generalized estimating equation models.
\newblock {\em Statistical Methods in Medical Research}, 29(6):1746--1762.

\bibitem[Robins, 1986]{robins1986new}
Robins, J. (1986).
\newblock A new approach to causal inference in mortality studies with a sustained exposure period--{Application} to control of the healthy worker survivor effect.
\newblock {\em Mathematical Modelling}, 7(9-12):1393--1512.

\bibitem[Robins et~al., 2016]{robins2016technical}
Robins, J., Li, L., {Tchetgen Tchetgen}, E., and {van der Vaart}, A. (2016).
\newblock Technical report: {Higher} order influence functions and minimax estimation of nonlinear functionals.
\newblock {\em arXiv preprint arXiv:1601.05820}.

\bibitem[Robins et~al., 2007]{robins2007comment}
Robins, J., Sued, M., Lei-Gomez, Q., and Rotnitzky, A. (2007).
\newblock Comment: {Performance} of double-robust estimators when “inverse probability” weights are highly variable.
\newblock {\em Statistical Science}, 22(4):544--559.

\bibitem[Robins et~al., 1994]{robins1994estimation}
Robins, J.~M., Rotnitzky, A., and Zhao, L.~P. (1994).
\newblock Estimation of regression coefficients when some regressors are not always observed.
\newblock {\em Journal of the American Statistical Association}, 89(427):846--866.

\bibitem[Robins and Wang, 2000]{robins2000inference}
Robins, J.~M. and Wang, N. (2000).
\newblock Inference for imputation estimators.
\newblock {\em Biometrika}, 87(1):113--124.

\bibitem[Rosenbaum, 1987]{rosenbaum1987model}
Rosenbaum, P.~R. (1987).
\newblock Model-based direct adjustment.
\newblock {\em Journal of the American Statistical Association}, 82(398):387--394.

\bibitem[Rosenblum and {van der Laan}, 2010]{rosenblum2010simple}
Rosenblum, M. and {van der Laan}, M.~J. (2010).
\newblock Simple, efficient estimators of treatment effects in randomized trials using generalized linear models to leverage baseline variables.
\newblock {\em The International Journal of Biostatistics}, 6(1):Article 13.

\bibitem[Seaman et~al., 2012]{seaman2012combining}
Seaman, S.~R., White, I.~R., Copas, A.~J., and Li, L. (2012).
\newblock Combining multiple imputation and inverse‐-probability weighting.
\newblock {\em Biometrics}, 68(1):129--137.

\bibitem[Stefanski and Boos, 2002]{stefanski2002the}
Stefanski, L.~A. and Boos, D.~D. (2002).
\newblock The calculus of {M-estimation}.
\newblock {\em The American Statistician}, 56(1):29--38.

\bibitem[Tackney et~al., 2023]{tackney2023}
Tackney, M.~S., Morris, T., White, I., Leyrat, C., Diaz-Ordaz, K., and Williamson, E. (2023).
\newblock A comparison of covariate adjustment approaches under model misspecification in individually randomized trials.
\newblock {\em Trials}, 24(1):14.

\bibitem[Tsiatis, 2006]{tsiatis2006the}
Tsiatis, A.~A. (2006).
\newblock The geometry of influence functions.
\newblock In {\em Semiparametric Theory and Missing Data}, pages 21--51. Springer New York.

\bibitem[Tsiatis et~al., 2008]{tsiatis2008covariate}
Tsiatis, A.~A., Davidian, M., Zhang, M., and Lu, X. (2008).
\newblock Covariate adjustment for two-sample treatment comparisons in randomized clinical trials: {A} principled yet flexible approach.
\newblock {\em Statistics in Medicine}, 27(23):4658--4677.

\bibitem[{US Food and Drug Administration}, 2023]{fda2023}
{US Food and Drug Administration} (2023).
\newblock Adjusting for covariates in randomized clinical trials for drugs and biological products.
\newblock \url{https://www.fda.gov/media/148910/download}.
\newblock Accessed Jun 1, 2024.

\bibitem[{van der Laan} and Rubin, 2006]{van2006targeted}
{van der Laan}, M.~J. and Rubin, D. (2006).
\newblock Targeted maximum likelihood learning.
\newblock {\em The International Journal of Biostatistics}, 2(1):Article 11.

\bibitem[{van Lancker} et~al., 2024]{van2024covariate}
{van Lancker}, K., Bretz, F., and Dukes, O. (2024).
\newblock Covariate adjustment in randomized controlled trials: {General} concepts and practical considerations.
\newblock {\em Clinical Trials}, 21(4):399--411.

\bibitem[Vansteelandt and Keiding, 2011]{vansteelandt2011invited}
Vansteelandt, S. and Keiding, N. (2011).
\newblock Invited commentary: {G-computation}--{Lost} in translation?
\newblock {\em American Journal of Epidemiology}, 173(7):739--742.

\bibitem[Vermeulen and Vansteelandt, 2015]{vermeulen2015bias}
Vermeulen, K. and Vansteelandt, S. (2015).
\newblock Bias-reduced doubly robust estimation.
\newblock {\em Journal of the American Statistical Association}, 110(511):1024--1036.

\bibitem[Wang et~al., 2023]{wang2023model}
Wang, B., Susukida, R., Mojtabai, R., Amin-Esmaeili, M., and Rosenblum, M. (2023).
\newblock Model-robust inference for clinical trials that improve precision by stratified randomization and covariate adjustment.
\newblock {\em Journal of the American Statistical Association}, 118(542):1152--1163.

\bibitem[Ye et~al., 2023a]{ye2023robust}
Ye, T., Bannick, M., Yi, Y., and Shao, J. (2023a).
\newblock Robust variance estimation for covariate-adjusted unconditional treatment effect in randomized clinical trials with binary outcomes.
\newblock {\em Statistical Theory and Related Fields}, 7(2):159--163.

\bibitem[Ye et~al., 2023b]{ye2023toward}
Ye, T., Shao, J., Yi, Y., and Zhao, Q. (2023b).
\newblock Toward better practice of covariate adjustment in analyzing randomized clinical trials.
\newblock {\em Journal of the American Statistical Association}, 118(544):2370--2382.

\bibitem[Yuan et~al., 2012]{yuan2012variable}
Yuan, S., Zhang, H.~H., and Davidian, M. (2012).
\newblock {Variable selection for covariate‐adjusted semiparametric inference in randomized clinical trials}.
\newblock {\em Statistics in Medicine}, 31(29):3789--3804.

\bibitem[Zhao et~al., 2024]{zhao2024hoif}
Zhao, S., Wang, X., Liu, L., and Zhang, X. (2024).
\newblock Covariate adjustment in randomized experiments motivated by higher-order influence functions.
\newblock {\em arXiv preprint arXiv:2411.08491}.

\end{thebibliography}

\clearpage

\appendix
\appendixpage
\beginappendix

\section{Working Model Specifications} \label{si:wm}

Without loss of generality, we do not consider any covariates transformed from $W_i$.  For the homogeneous working model, there is no interaction between $A_i$ and $W_i$.  Then, $\bbeta^\top X_i$ can be written as 
\[
    \beta_1^A I(A_i=1) + \beta_2^A I(A_i=2) + \bbeta_{0,W}^\top W_i.
\]
On the other hand, the heterogeneous working model includes all those interaction terms.  Then, $\bbeta^\top X_i$ can be written as 
\begin{multline*}
    I(A_i=1)(\beta_1^A + \bbeta_{1,W}^\top W_i) + I(A_i=2)(\beta_2^A + \bbeta_{2,W}^\top W_i) = \beta_1^A I(A_i=1) + \beta_2^A I(A_i=2) + \bbeta_{0,W}^\top W_i + \\  \{ \bbeta_{1,W} - \bbeta_{0,W} \}^\top W_i \times I(A_i=1) + \{ \bbeta_{2,W} - \bbeta_{0,W} \}^\top W_i \times I(A_i=2).
\end{multline*}

\section{Variance Estimator proposed in \texorpdfstring{\citet{ye2023robust}}{Ye et al. (2023a)}} \label{si:var-ye}  

Let $\uSigma_{\bmu} ( a, b )$ be the $( a, b )$-cell of $\uSigma_{\bmu}$.  The variance estimator proposed by \citet{ye2023robust} is
\[
    \uSigma_{\bmu} ( a, b ) = \ 
    \begin{cases}
        \begin{aligned} \frac{1}{n \pi_a} \hat{\Var} \left[ Y_i - m ( \hbbeta^\top X_i ) \middle\vert A_i = a \right] & + \frac{2}{n} \hat{\Cov} \left[ Y_i, m( \hbbeta^\top X_i ) \middle\vert A_i = a \right] \\ & - \frac{1}{n} \hat{\Var} \left[ m ( \hbbeta^\top X_{i ( a )} ) \right] \end{aligned} & a = b,  \\
        \begin{aligned} \frac{1}{n} \hat{\Cov} \left[ Y_i, m ( \hbbeta^\top X_{i(b)} ) \middle\vert A_i = a \right] + \frac{1}{n} \hat{\Cov} \left[ Y_i, m ( \hbbeta^\top X_{i(a)} ) \middle\vert A_i = b \right] \\ - \frac{1}{n} \hat{\Cov} \left[ m ( \hbbeta^\top  X_{i ( a )} ), m ( \hbbeta^\top X_{i ( b )} ) \right] \end{aligned}  & a \neq b.
    \end{cases} 
\]

\section{Proposed Score-based Method for Ratios} \label{si:ratio}

For $H_R: \mu_2 /\mu_1 = \delta_0$, the generalized score statistic is (e.g., $\mu_1 > 0, \mu_2 \geq 0$)
\begin{equation}
    \hQ_R = \frac{ ( \hmu_2 - \delta_0 \hmu_1 )^2 }{ \hSigma_{\bmu}(2,2) - 2\delta_0\hSigma_{\bmu}(2,1) + \delta_0^2\hSigma_{\bmu}(1,1) + ( \hmu_2 - \delta_0\hmu_1 )^2 / n } \label{eq:score-stats-ratio}.
\end{equation}
By inverting the hypothesis test, the corresponding confidence interval can be obtained by solving the quadratic equation $\delta_0^2 - 2 a (\hmu_2/\hmu_1) \delta_0 + b (\hmu_2/\hmu_1)^2 = 0 $, where 
\begin{align*}
    a & = \ \frac{ 1 - \chi^2_{1-\alpha} \times \left\{ \hSigma_{\bmu}(2,1) / (\hmu_1\hmu_2) + 1/n \right\} }{ 1 - \chi^2_{1-\alpha} \times \left\{ \hSigma_{\bmu}(1,1) / \hmu_1^2 + 1/n \right\} } \\
    b & = \ \frac{ 1 - \chi^2_{1-\alpha} \times \left\{ \hSigma_{\bmu}(2,2) / \hmu_2^2 + 1/n \right\} }{ 1 - \chi^2_{1-\alpha} \times \left\{ \hSigma_{\bmu}(1,1) / \hmu_1^2 + 1/n \right\} }.
\end{align*}
Assume that $( 1 - \chi^2_{1-\alpha} / n ) \times \hmu_1^2 > \chi^2_{1-\alpha} \times \hSigma_{\bmu}(1,1)$, which generally holds when $\hmu_1$ is not too close to zero.  When $a^2 - b > 0$, the $(1-\alpha)\times100\%$ score interval is then 
\begin{equation}
     \frac{\mu_2}{\mu_1} \in  \left[ \frac{\hmu_2}{\hmu_1} \times ( a - \sqrt{a^2-b} ) , \frac{\hmu_2}{\hmu_1} \times ( a + \sqrt{a^2-b} ) \right]. \label{eq:score-ci-ratio}  
\end{equation} 
When $n$ is not too small, $a^2 - b > 0$ generally holds as 
\begin{align*}
    & \left[ 1 - \chi^2_{1-\alpha} \times \left\{ \hSigma_{\bmu}(1,1) / \hmu_1^2 + 1/n \right\} \right]^2 ( a^2 - b ) \\
    = & \ \left\{ \chi^2_{1-\alpha} - \left( \chi^2_{1-\alpha} \right)^2 / n \right\} \times \left\{ \hSigma_{\bmu}(2,2) / \hmu_2^2 - 2\hSigma_{\bmu}(2,1)  / (\hmu_1\hmu_2) + \hSigma_{\bmu}(1,1) / \hmu_1^2 \right\} \\ 
    & \quad\quad + \left( \chi^2_{1-\alpha} \right)^2 \times \left\{ \left( \hSigma_{\bmu}(2,1)  / (\hmu_1\hmu_2) \right)^2 - \left( \hSigma_{\bmu}(1,1) / \hmu_1^2 \right) \left( \hSigma_{\bmu}(2,2) / \hmu_2^2 \right) \right\} \\
    = & \left\{ \chi^2_{1-\alpha} - \left( \chi^2_{1-\alpha} \right)^2 / n \right\} \times \hat{\Var} \left[ \log(\hmu_2/\hmu_1) \right] \\
    & \quad\quad + \left( \chi^2_{1-\alpha} \right)^2 \times \left\{ \left( \hat{\Cov} \left[ \log\hmu_2, \log\hmu_1 \right] \right)^2 - \hat{\Var} \left[ \log\hmu_1 \right] \times \hat{\Var} \left[ \log\hmu_2 \right] \right\} \\
    = & \ \chi^2_{1-\alpha} \times \hat{\Var} \left[ \log(\hmu_2/\hmu_1) \right] + O \left( 1 / n^2 \right).
\end{align*}
The second term in the above equation can be ignored as $\hat{\Var} \left[ \log(\hmu_2/\hmu_1) \right] = O(1/n)$.  Besides, 
\begin{align*}
    & \log \left( a \pm \sqrt{a^2-b} \right) \\
    = & \ \log \left( \frac{ 1 - \chi^2_{1-\alpha} \times \left\{ \hSigma_{\bmu}(2,1) / (\hmu_1\hmu_2) + 1/n \right\} \pm \sqrt{\chi^2_{1-\alpha} \times \hat{\Var} \left[ \log(\hmu_2/\hmu_1) \right] + O ( 1/n^2 ) } }{ 1 - \chi^2_{1-\alpha} \times \left\{ \hSigma_{\bmu}(1,1) / \hmu_1^2 + 1/n \right\} } \right) \\
    = & \ \log \left( \frac{ 1 - O (1/n) \pm \sqrt{ \chi^2_{1-\alpha} \times \hat{\Var} \left[ \log(\hmu_2/\hmu_1) \right] + O ( 1/n^2 ) } }{ 1 - O (1/n) } \right) \\
    = & \ \pm \frac{ \sqrt{\chi^2_{1-\alpha} \times \hat{\Var} \left[ \log(\hmu_2/\hmu_1) \right] + O ( 1/n^2 ) } }{ 1 - O (1/n) } + O ( 1 / n ),
\end{align*}
since $\sqrt{\hat{\Var} \left[ \log(\hmu_2/\hmu_1) \right] } = O (1/\sqrt{n})$.  This implies that the log-scale of the interval \eqref{eq:score-ci-ratio}, when $n$ is large, is approximately symmetric about $\log\hmu_2 - \log\hmu_1$, and it is almost the same as the Wald interval for $\log\hmu_2 - \log\hmu_1$.  

Analogous to the proposed method for $H_D$ discussed in Section~\ref{sec:method} in the main manuscript, the variance estimator $\hSigma_{\bmu}$ in \eqref{eq:score-stats-ratio} and \eqref{eq:score-ci-ratio} can be replaced with either $\tSigma_{\bmu}$ or $\uSigma_{\bmu}$.  For stratified/biased-coin randomization, the statistical inference based on $\hQ_R$ is conservative, hence asymptotically valid, under these two randomization schemes. 

\section{Simulation Specifications}  \label{si:sim-specs}

For those scenarios of one baseline covariate with $\beta_W = \log(2)$, and three baseline covariates with $\beta_j^W=\sqrt{\log(2)^2/3}$, we have those values for $\beta_1^A$ and $\beta_2^A$ as listed below:
\begin{itemize}
    \item $( -0.9355, -0.2224 )$ for $N = 326$, $\mu_1 =30 \%$ and $\mu_2 = 45 \%$;
    \item $( -3.1555, -1.9878 )$ for $N = 326$, $\mu_1 = 5 \%$ and $\mu_2 = 13.9 \%$;
    \item $( -0.9355, 0.4492 )$ for $N = 80$, $\mu_1 = 30 \%$ and $\mu_2 = 60 \%$.
\end{itemize}
For those scenarios of one baseline covariate with $\beta_W = \log(1.1)$, and three baseline covariates with $\beta_j^W=\sqrt{\log(1.1)^2/3}$, we have those values for $\beta_1^A$ and $\beta_2^A$ as listed below:
\begin{itemize}
    \item $( -0.8491, -0.2011 )$ for $N = 326$, $\mu_1 =30 \%$ and $\mu_2 = 45 \%$;
    \item $( -2.9485, -1.8269 )$ for $N = 326$, $\mu_1 = 5 \%$ and $\mu_2 = 13.9 \%$;
    \item $( -0.8491, 0.4064 )$ for $N = 80$, $\mu_1 = 30 \%$ and $\mu_2 = 60 \%$.
\end{itemize}

\section{Additional Simulations}

\subsection{Simulation studies for variance estimators} \label{si:sim-var}

\begin{figure}[ht!]
    \centering
    \includegraphics[width=\linewidth]{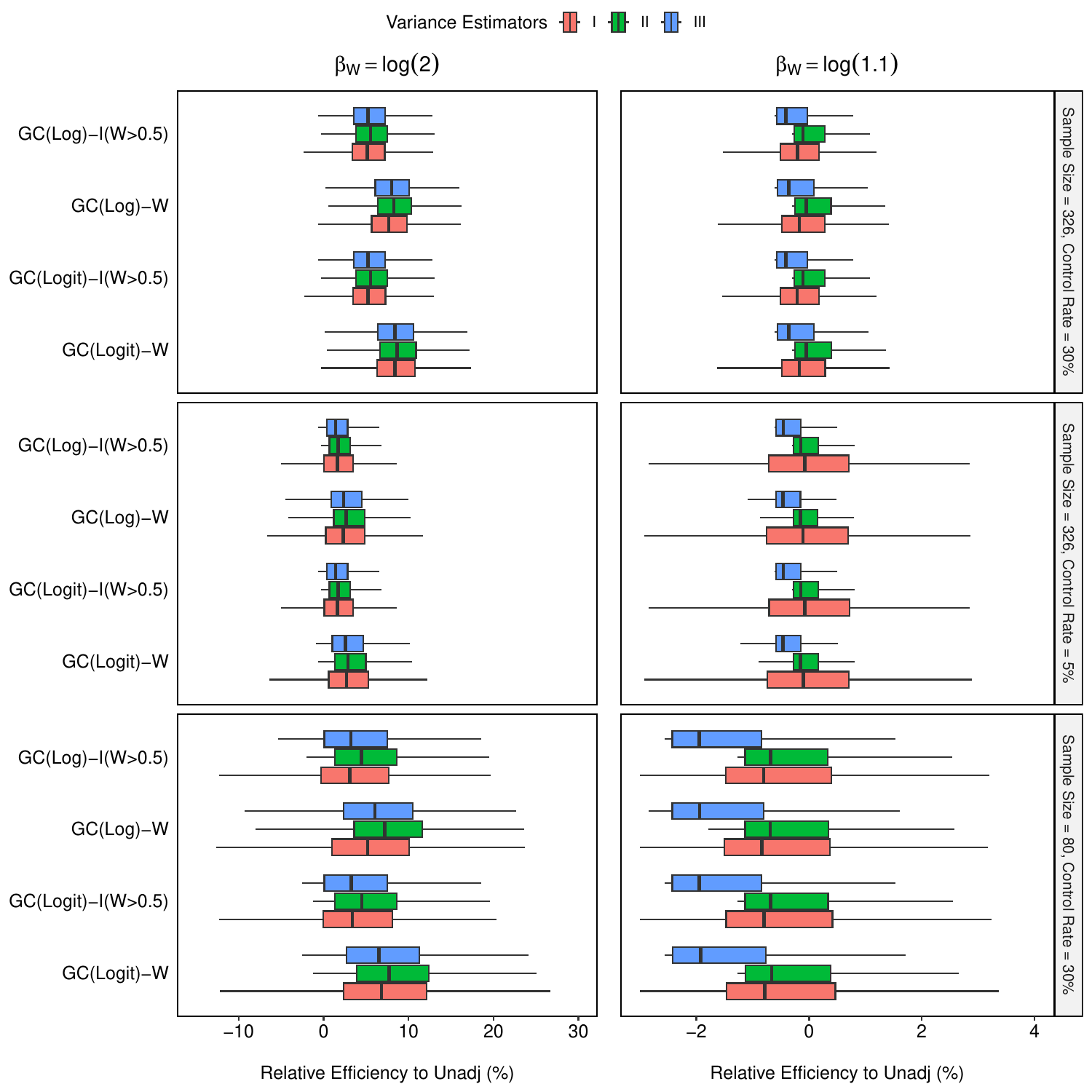}
    \caption{The efficiency of g-computation estimators using the three variance estimators, $\hSigma_{\bmu}$ (I), $\tSigma_{\bmu}$ (II), and $\uSigma_{\bmu}$ (III) in Table~\ref{tab:var} of the main manuscript.  The relative efficiency is calculated for each simulated trial as one minus the ratio of the estimated variance between the g-computation and the unadjusted estimators.} 
    \label{fig:re}
\end{figure}

We generate binary outcomes from a Bernoulli distribution with $\pr ( Y_i=1 \vert A_i=a, W_i ) = \text{expit}(\beta_a^A + \beta_W W_i)$, with one baseline covariate ($W_i$) from a standard normal distribution.  We simulate the three hypothetical trials described in Section~\ref{sec:sim-small} in the main manuscript, and for each trial, we consider two types of $(Y, W)$-association, $\beta_W = \log(1.1), \log(2)$.  To keep the sample sizes of two arms exactly the same, the completely randomized design is used.  Those values for $(\beta_W,\beta_1^A,\beta_2^A)$ are provided in Section \ref{si:sim-specs}.  Two g-computation estimators for $\mu_2-\mu_1$ are evaluated, denoted by \textbf{GC(Logit)} and \textbf{GC(Log)}.  The former one uses logistic regression, while the latter one uses Poisson regression (which is included to further explore the performance under model misspecification).  The unadjusted (\textbf{Unadj}) estimator is also included for the comparisons.  The two g-computation estimators adjust for one baseline covariate of either $W_i$ or $I(W_i>0.5)$.  We evaluate the three existing variance estimators in Table~\ref{tab:var} of the main manuscript, and summarize their relative efficiency in Figure~\ref{fig:re}.

\clearpage

\subsection{Simulation results for interval coverage probabilities} \label{si:sim-cp}

\begin{figure}[ht!]
    \centering
    \includegraphics[width=\linewidth]{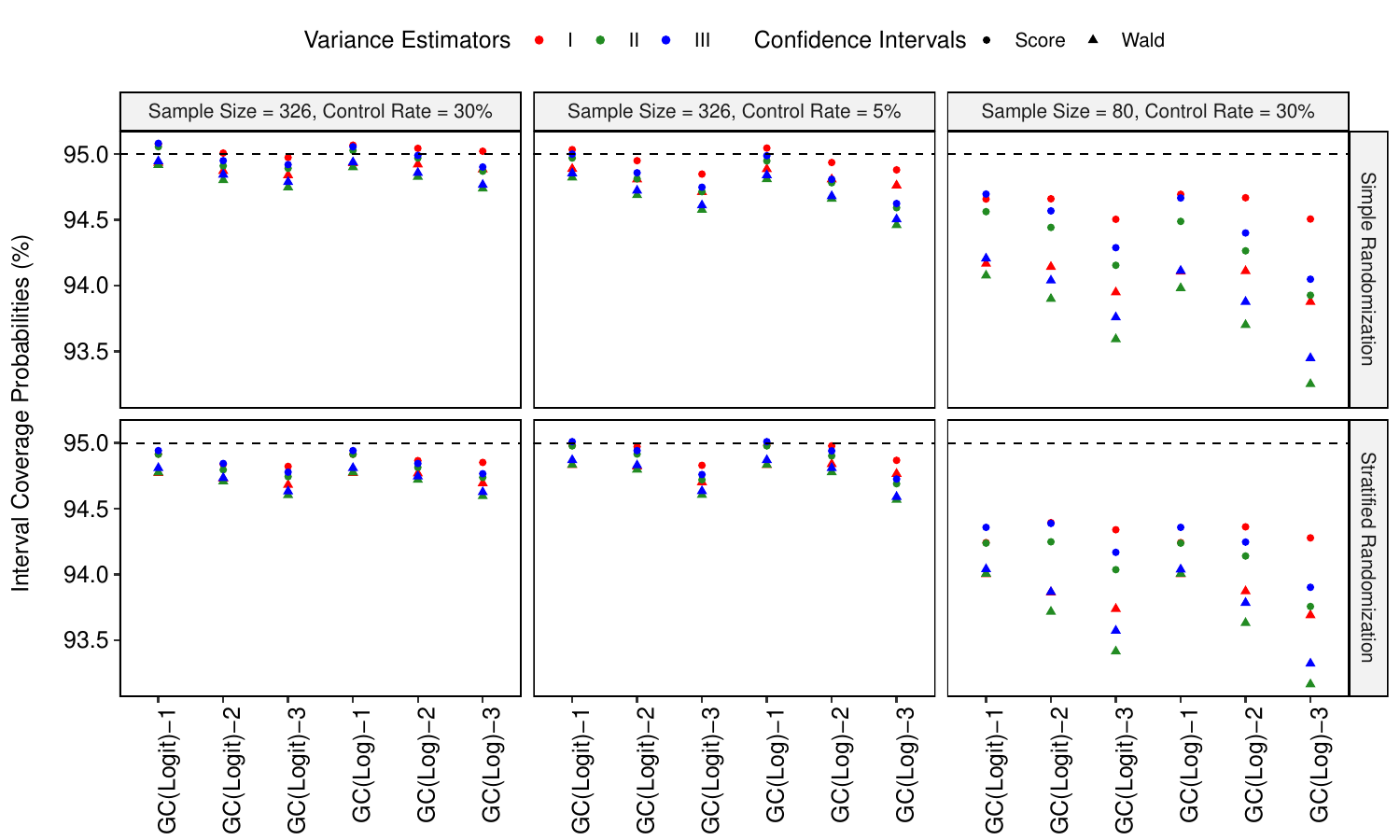}
    \caption{The coverage probability of the 95\% Wald interval and the proposed score interval under the alternative, evaluated for two g-computation estimators (of $\mu_2-\mu_1$), using logistic (GC(Logit)) and Poisson (GC(Log)) regression models, respectively, under three hypothetical trials (by column) and two randomization schemes (by row).  For simple randomization, both g-computation estimators adjust for either $W_{i1}$ (1), $(W_{i1},W_{i2})$ (2), or $(W_{i1},W_{i2},W_{i3})$ (3).  For stratified randomization, they adjust for either $S_{i}$ (1), $(S_i,W_{i1})$ (2), or $(S_i,W_{i1},W_{i2})$ (3).  All three variance estimators, $\hSigma_{\bmu}$ (I), $\tSigma_{\bmu}$ (II), and $\uSigma_{\bmu}$ (III), summarized in Table~\ref{tab:var} in the main manuscript, are used in the two types of confidence intervals. }
\end{figure}

\clearpage

\subsection{Simulation results for potentially less biased variance estimators} \label{si:sim-bc-var}

\begin{figure}[ht!]
    \centering
    \includegraphics[width=\linewidth]{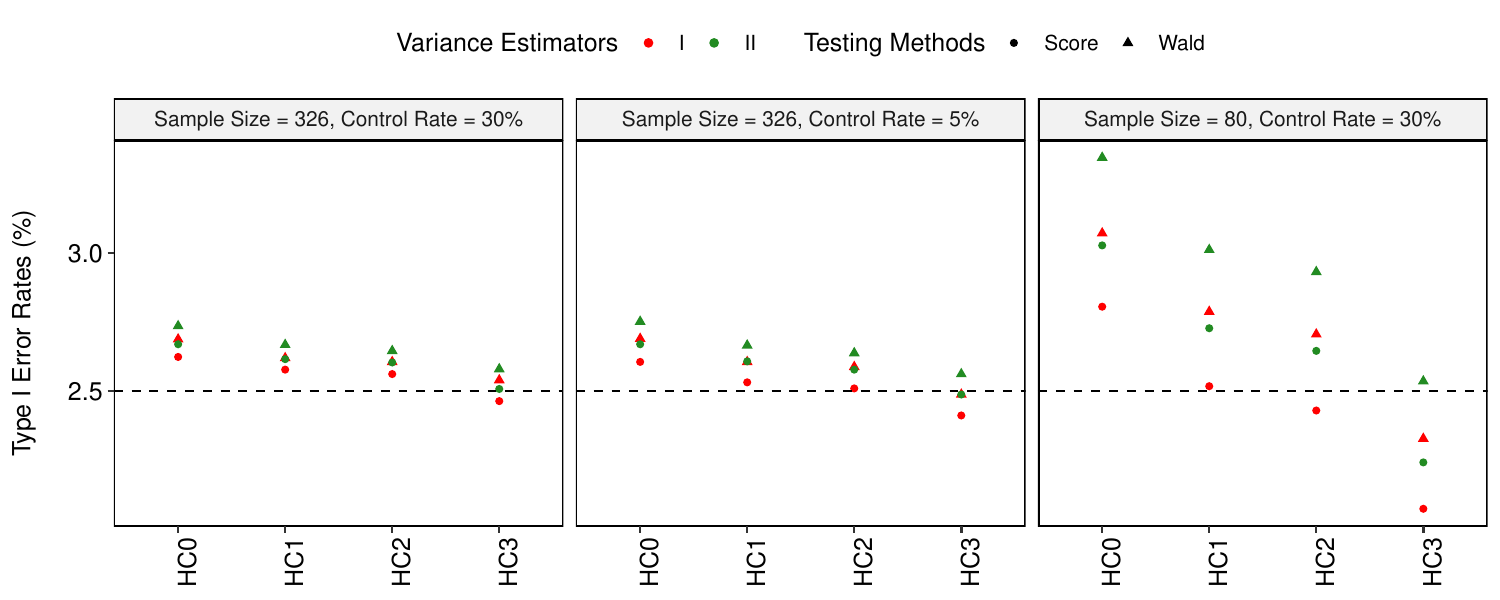}
    \caption{ The type I error rate for testing $H_0:\mu_2-\mu_1=0$ using the Wald test and the proposed score test for g-computation estimators using potentially less biased variance estimators.  The logistic regression model is used to adjust for $(W_{i1},W_{i2},W_{i3})$ under the three hypothetical trials in Section~\ref{sec:sim-small} in the main manuscript.  Two variance estimators, $\hSigma_{\bmu}$ (I) and $\tSigma_{\bmu}$ (II) from Table~\ref{tab:var}, with/without correction are used in the two testing methods: HC0 = no correction; HC1 = correction for degree-of-freedom \citep{tsiatis2008covariate}; HC2 = \citet{kauermann2001note}; HC3 = \citet{mancl2001covariance}.}
\end{figure}

\clearpage
\FloatBarrier

\section{Proofs}

\subsection{``Score-based'' influence functions} \label{si:eif-mu-mest}

We show that the influence function of $\hbmu$ under simple randomization is
\[
    \bpsi_{\bmu}(D_i) = G_{\bbeta} \bpsi_{\bbeta}(D_i) + \bg(W_i;\bbeta_0) - \bmu
\]
where $\bg(W_i;\bbeta) = \left( m(\bbeta^\top X_{i(1)}), m(\bbeta^\top X_{i(2)}) \right)^\top$ and $G_{\bbeta} = \E\big[\deriv{\bg(W_i;\bbeta_0)}{\bbeta}\big]$. Write $B_{\btheta}$ as  
\[
    \begin{pmatrix}
        \mathbf{I}_{2 \times 2} & -G_{\beta} \\
        \mathbf{0}_{p \times 2} & B_{\bbeta}
    \end{pmatrix},
\]
where $p$ is the dimension of $\bbeta$, and thus 
\[
    B_{\btheta}^{-1} = 
    \begin{pmatrix}
        \mathbf{I}_{2 \times 2} & G_{\beta}B_{\bbeta}^{-1} \\
        \mathbf{0}_{p \times 2} & B_{\bbeta}^{-1}
    \end{pmatrix}.
\]
Next, write $M_{\btheta}$ as 
\[
    \begin{pmatrix}
        M_{\bmu\bmu} &  M_{\bmu\bbeta} \\
        M_{\bmu\bbeta}^\top & M_{\bbeta\bbeta}
    \end{pmatrix},
\]
where 
\begin{align*}
    M_{\bmu\bmu} &= \E\left[\left\{\bg(W_i;\bbeta_0) - \bmu\right\}\left\{\bg(W_i;\bbeta_0) - \bmu\right\}^\top\right], \\
    M_{\bmu\bbeta} &= \E\left[\left\{\bg(W_i;\bbeta_0) - \bmu\right\}X_i^\top\left\{Y_i-m(\bbeta_0 X_i)\right\}\right], \\
    M_{\bbeta\bbeta} &=  M_{\bbeta}.
\end{align*}
Above all, the first $2 \times 2$ block of $B_{\btheta}^{-1} M_{\btheta} \{ B_{\btheta}^{-1} \}^\top$ is
\[
   \left\{G_{\bbeta}\bpsi_{\bbeta} + \bg(W_i;\bbeta_0) - \bmu\right\}\left\{G_{\bbeta}\bpsi_{\bbeta} + \bg(W_i;\bbeta_0) - \bmu\right\}^\top,
\]
which implies that $\Sigma_{\bmu} = \E \left\{ \bpsi^\score_{\bmu}(D_i) \bpsi_{\bmu}^\score(D_i)^\top \right\} / n$. Thus, $\bpsi^\score_{\bmu}(D_i)$ is the influence function for the partial $M$-estimator $\hbmu$.

\subsection{Proof of Lemma \ref{lem:strata}} \label{si:strata}

By Bayes' rule and iterative expectation, we have, for any $s, a$
\begin{align*}
\pr (S_i = s \vert A_i= a) & = \ \pr (S_i = s) \pr (A_i = a \vert S_i = s) / \pr (A_i = a) \\
& = \ \pr (S_i = s) \pi_a / \pi_a = \pr (S_i = s). 
\end{align*}

\subsection{Proof of Proposition~\ref{thm:eif-mu}} \label{si:eif-mu-semiparam}

Let $X_i^\top = (J_i^\top, Z_i^\top)$ and $X_{i(a)}^{\top} = (J(a)^\top, Z_{i(a)}^{\top})$, where $J_i=(I(A_i=1),I(A_i=2))^\top$ and $J(a)$ is a two-dimensional column vector with only the $a$-th row being 1 and the other being 0. Hence $J_i^{\top} J (a) \equiv I (A_i = a)$. Write 
\[
B_{\bbeta} = \E \left[ m^\prime(\bbeta^\top_0 X_i) X_i X_i^\top \right ] = 
    \begin{pmatrix}
        \E \left[ m^\prime(\bbeta^\top_0 X_i) J_i J_i^\top \right] & \E \left[ m^\prime(\bbeta^\top_0 X_i) J_i Z_i^\top \right]\\ 
        \E \left[ m^\prime(\bbeta^\top_0 X_i) Z_i J_i^\top \right] & \E \left[ m^\prime(\bbeta^\top_0 X_i) Z_i Z_i^\top \right]
    \end{pmatrix}.
\]
Note that $J(a)$ is not a random vector, so we can obtain the following:
\[
\begin{split}
    B_{\bbeta} \cdot \begin{pmatrix} J(a) / \pi_a \\ \bzero \end{pmatrix}
    = & \begin{pmatrix} 
        \E \left[ m^\prime(\bbeta^\top_0 X_i) J_i I (A_i = a) \right] / \pi_a \\ 
        \E \left[ m^\prime(\bbeta^\top_0 X_i) Z_i I (A_i = a) \right] / \pi_a \end{pmatrix} \\
    = & \begin{pmatrix}  
        \E \left[ \left. m^\prime(\bbeta^\top_0 X_{i(a)}) J(a) \right\vert A_i=a \right] \\ 
        \E \left[ \left. m^\prime(\bbeta^\top_0 X_{i(a)}) Z_{i(a)} \right\vert A_i=a \right] \end{pmatrix} \\
    \equiv & \ \E\left[ \left. m^\prime(\bbeta^\top_0 X_{i(a)}) X_{i(a)} \right\vert A_i=a \right]. 
\end{split}
\]
For simple randomization, the last term in the above equals to $\E \left[m^\prime(\bbeta^\top_0 X_{i(a)})X_{i(a)}\right]$.  In the following, we show that this equality also holds for stratified/biased-coin randomization.  

As $m^\prime(\bbeta^\top_0 X_{i(a)})X_{i(a)}$ only depends on $W_i$, $A_i \perp W_i \vert S_i$, and $\pr (S_i = s \vert A_i = a) = \pr (S_i = s)$ for any $s$ (Lemma~\ref{lem:strata}), we have that 
\[
\begin{split}
    \E \left[ \left. m^\prime(\bbeta^\top_0 X_{i(a)})X_{i(a)} \right\vert A_i = a \right] 
    = & \sum_{s}\E \left[ \left. m^\prime(\bbeta^\top_0 X_{i(a)})X_{i(a)} \right\vert A_i = a, S_i = s\right] \pr (S_i = s \vert A_i=a)  \\
    = & \sum_{s} \E \left[ \left. m^\prime(\bbeta^\top_0 X_{i(a)})X_{i(a)} \right\vert S_i = s\right] \pr (S_i = s)\\ 
    = & \ \E \left[m^\prime(\bbeta^\top_0 X_{i(a)})X_{i(a)}\right].
\end{split}  
\]

Combining the above calculations, we have the following important identity
\begin{align*}
\E \left[ m^\prime(\bbeta^\top_0 X_{i(a)}) X_{i(a)} \right]^{\top} = \begin{pmatrix} J(a)^\top/\pi_a & \bzero^\top \end{pmatrix} B_{\bbeta}
\end{align*}
which further implies
\begin{align*}
    \E \left[ m^\prime(\bbeta^\top_0 X_{i(a)}) X_{i(a)} \right]^{\top} B_{\bbeta}^{-1} \equiv \begin{pmatrix} J(a)^\top/\pi_a & \bzero^\top \end{pmatrix}.
\end{align*}
Therefore, 
\[
\begin{split}
    \E\left[m^\prime(\bbeta_0^\top X_{i(a)})X_{i(a)}\right]^\top \bpsi_{\bbeta}(D_i) 
    = & \ \E\left[m^\prime(\bbeta_0^\top X_{i(a)})X_{i(a)}\right]^\top B_{\bbeta}^{-1}X_i\left\{Y_i-m(\bbeta_0^\top X_i)\right\} \\
    = & \begin{pmatrix} J (a)^\top/\pi_a & \bzero^\top \end{pmatrix} \cdot \begin{pmatrix} J_i \\ Z_i \end{pmatrix} \left\{Y_i-m(\bbeta_0^\top X_i)\right\} \\
    = & \ \frac{I(A_i=a)}{\pi_a}\left\{Y_i-m(\bbeta_0^\top X_i)\right\}.
\end{split}
\]
Replacing the first term in (\ref{eq:if-mu}) with the above term, we conclude that $\psi_{a}^\score (D_i)$ has the same form $\psi_{a}^\aipw (D_i)$.

\subsection{Proof of Proposition~\ref{thm:var-dec}} \label{si:avar-ye}

We prove the result by applying Proposition~\ref{thm:eif-mu} and Lemma~\ref{lem:strata} to the decomposition of $\Sigma_{\bmu}$ discussed in Section~\ref{sec:var-dec} in the main manuscript. We first write such decomposition as 
\[
    \Sigma_{\bmu} = \Sigma_m + \Sigma_w + \Sigma_{m,w} + \Sigma_{m,w}^\top.
\]
Applying Proposition~\ref{thm:eif-mu}, we have
\begin{align*}
    n\Sigma_m(a,a) = & \ \pi_a^{-1} \Var \left[ \left. Y_i(a) - m(\bbeta_0^\top X_{i(a)}) \right\vert A_i=a \right], \\
    n\Sigma_m(a,b) = & \ 0, \\
    n\Sigma_{m,w}(a,b) = & \ \Cov \left[ \left. Y_i(a), m(\bbeta_0^\top X_{i(b)}) \right\vert A_i=a \right] - \Cov \left[ \left. m(\bbeta_0^\top X_{i(a)}), m(\bbeta_0^\top X_{i(b)}) \right\vert A_i=a \right].
\end{align*}

For simple randomization, it is straightforward to see that 
\begin{align}
    \Var \left[ \left. Y_i(a) - m(\bbeta_0^\top X_{i(a)}) \right\vert A_i=a \right] = & \ \Var \left[ Y_i(a) - m(\bbeta_0^\top X_{i(a)}) \right], \label{eq:ye-var} \\
    \Cov \left[ \left. Y_i(a), m(\bbeta_0^\top X_{i(b)}) \right\vert A_i=a \right] = & \ \Cov \left[ Y_i(a), m(\bbeta_0^\top X_{i(b)}) \right], \label{eq:ye-cov1} \\
    \Cov \left[ \left. m(\bbeta_0^\top X_{i(a)}), m(\bbeta_0^\top X_{i(b)}) \right\vert A_i=a \right] = & \ \Cov \left[ m(\bbeta_0^\top X_{i(a)}), m(\bbeta_0^\top X_{i(b)}) \right]. \label{eq:ye-cov2}
\end{align}
Note that the second component ($\Sigma_{w}$) remains the same. Then, 
\begin{align*}
    n\Sigma_{\bmu}(a,a) = & \ \pi_a^{-1} \Var \left[ Y_i(a) - m(\bbeta_0^\top X_{i(a)}) \right] + 2 \Cov \left[ Y_i(a), m(\bbeta_0^\top X_{i(a)} ) \right] \\ & - \Var \left[ m(\bbeta_0^\top X_{i(a)}) \right], \\
    n\Sigma_{\bmu}(a,b) = & \ \Cov \left[Y_i(a), m(\bbeta_0^\top X_{i(b)}) \right] + \Cov \left[ Y_i(b), m(\bbeta_0^\top X_{i(a)}) \right] \\ & - \Cov \left[ m(\bbeta_0^\top X_{i(a)}), m(\bbeta_0^\top X_{i(b)}) \right],
\end{align*}
which indicates that $n\Sigma_{\bmu}$ is identical to the asymptotic covariance of $\sqrt{n}(\hbmu-\bmu)$ proposed by \citet{ye2023robust}. 

Next, we show \eqref{eq:ye-var}, \eqref{eq:ye-cov1} and \eqref{eq:ye-cov2} hold under stratified (and biased-coin) randomization. As $g^\prime(\bbeta^\top_0 X_{i(a)})X_{i(a)}$ only depends on $W_i$, $A_i \perp W_i \vert S_i$, and $p(S_i \vert A_i=a)=p(S_i)$ (Lemma~\ref{lem:strata}), we prove \eqref{eq:ye-var} via the law of total variance as follows:
\[
\begin{split}
      & \E \left\{ \left. \Var \left[ \left. Y_i(a) - m(\bbeta_0^\top X_{i(a)}) \right\vert S_i, A_i=a \right] \right\vert A_i=a \right\} \\
    = & \sum_{s} \Var \left[ \left. Y_i(a) - m(\bbeta_0^\top X_{i(a)}) \right\vert S_i=s, A_i=a \right] p(S_i=s \vert A_i=a) \\
    = & \sum_{s} \Var \left[ \left. Y_i(a) - m(\bbeta_0^\top X_{i(a)}) \right\vert S_i=s \right] p(S_i=s)  \\
    = &\ \E \left\{ \Var \left[ \left. Y_i(a) - m(\bbeta_0^\top X_{i(a)}) \right\vert S_i \right] \right\}, 
\end{split}
\]
\begin{align}
      & \Var \left\{ \left. \E \left[ \left. Y_i(a) - m(\bbeta_0^\top X_{i(a)}) \right\vert S_i, A_i=a \right] \right\vert A_i=a \right\} \nonumber\\
    = & \sum_{s} \left\{ \E \left [ \left. Y_i(a) - m(\bbeta_0^\top X_{i(a)}) \right\vert S_i=s, A_i=a \right] \right\}^2 p(S_i=s \vert A_i=a) \nonumber \\
    = & \sum_{s} \left\{ \E \left[ \left. Y_i(a) - m(\bbeta_0^\top X_{i(a)}) \right]^2 \right\vert S_i=s \right\}^2 p(S_i=s) \nonumber \\
    = & \ \Var \left\{ \E \left[ \left. Y_i(a) - m(\bbeta_0^\top X_{i(a)}) \right\vert S_i \right] \right\}. \label{eq:ye-step}
\end{align}
Note that $\E[Y_i(a) \vert A_i] = \mu_a$. We prove \eqref{eq:ye-cov1} as follows:
\[
\begin{split}
        & \Cov \left[ Y_i(a), m(\bbeta_0^\top X_{i(b)}) \right] \\
      = & \ \E \left\{ \Cov \left[ \left. Y_i(a), m(\bbeta_0^\top X_{i(b)}) \right\vert A_i \right] \right\} + \Cov \left\{ \E \left[ Y_i(a) \vert A_i \right], \E \left[ m(\bbeta_0^\top X_{i(b)}) \vert A_i \right] \right\} \\
      = & \ \E \left\{ \Cov \left[ \left. Y_i(a), m(\bbeta_0^\top X_{i(b)}) \right\vert A_i \right] \right\} \\ 
      = & \sum_{c=1}^{2} \pi_c \Cov \left[ \left. Y_i(a), m(\bbeta_0^\top X_{i(b)}) \right\vert A_i=c \right] \quad\quad\text{following the proof for \eqref{eq:ye-step}} \\
      = & \sum_{c=1}^{2} \pi_c \Cov \left[ Y_i(a), m(\bbeta_0^\top X_{i(b)}) \right] = \Cov \left[ Y_i(a), m(\bbeta_0^\top X_{i(b)}) \right].
\end{split}
\]
Note that $\E[m(\bbeta_0^\top X_{i(a)}) \vert A_i] = \E[m(\bbeta_0^\top X_{i(a)})]$ (Section~\ref{si:eif-mu-semiparam}). We prove \eqref{eq:ye-cov2} as follows
\[
\begin{split}
        & \Cov[m(\bbeta_0^\top X_{i(a)}), m(\bbeta_0^\top X_{i(b)})] \\
      = & \ \E \left\{ \Cov \left[ \left. m(\bbeta_0^\top X_{i(a)}), m(\bbeta_0^\top X_{i(b)}) \right\vert A_i \right] \right\} + \Cov \left\{ \E \left[ \left. m(\bbeta_0^\top X_{i(a)}) \right\vert A_i \right], \E \left[ \left. m(\bbeta_0^\top X_{i(b)}) \right\vert A_i \right] \right\} \\
      = & \ \E \left\{ \Cov \left[ m(\bbeta_0^\top X_{i(a)}), m(\bbeta_0^\top X_{i(b)}) \vert A_i \right] \right\} \\ 
      = & \sum_{c=1}^{2} \pi_c \Cov \left[ m(\bbeta_0^\top X_{i(a)}), m(\bbeta_0^\top X_{i(b)}) \vert A_i=c \right] \quad\quad\text{following the proof for \eqref{eq:ye-step}} \\
      = & \sum_{c=1}^{2} \pi_c \Cov \left[ m(\bbeta_0^\top X_{i(a)}), m(\bbeta_0^\top X_{i(b)}) \right] = \Cov \left[ m(\bbeta_0^\top X_{i(a)}), m(\bbeta_0^\top X_{i(b)}) \right].
\end{split}
\]

\subsection{Generalized score statistics under simple randomization} \label{si:gscore-stats-sr}

\textbf{Difference in means.} Let $\ES U(D_i)$ be the stacked estimating equations for $\hbtheta = ( \hbmu, \hbbeta )$, and $h(\btheta) = \mu_2 - \mu_1 - \delta_0$. Let $(\bmu^\ast,\bbeta^\ast)$ be the solution of $\ES U(D_i) - \lambda H^{\ast\top}=0$ under the constraint $\mu^\ast_2 - \mu^\ast_1 = \delta_0$, where 
\[
    H^\ast = \deriv{h(\btheta^\ast)}{\btheta^\top} = ( J(2)^\top - J(1)^\top, \bzero_p^\top ).
\]
Then, $\bbeta^\ast = \hbbeta$ and
\begin{align*}
        \ES \left\{ m(\hbbeta^\top X_{i(1)}) - \mu^\ast_1 \right\} + \lambda = & \ 0, \\
        \ES \left\{ m(\hbbeta^\top X_{i(2)}) - \mu^\ast_2 \right\} - \lambda = & \ 0.
\end{align*}
Thus, $\lambda = n(\hmu_2 - \hmu_1 - \delta_0) / 2$. Under simple randomization, the generalized score statistic is defined as 
\begin{equation}
    n^{-1} \left\{ \ES U^\ast(D_i) \right\}^\top \left( H^\ast B_{\btheta}^{\star -1} \right)^\top \left[ H^\ast B_{\btheta}^{\star -1} M_{\btheta}^\ast \left( H^\ast B_{\btheta}^{\star -1} \right)^\top \right]^{-1} H^\ast B_{\btheta}^{\star -1} \left\{ \ES U^\ast(D_i) \right\} \label{eq:gs-stats}
\end{equation}
where $(B_{\btheta}^\ast, M_{\btheta}^\ast, U^\ast(D_i))$ are estimates of $(B_{\btheta}, M_{\btheta}, U(D_i))$ under $H_0$.  Note that $B_{\btheta}$ only depends on $\bbeta$. Then
\[
    B_{\btheta}^\star = \hB_{\btheta}=
    \begin{pmatrix}
        \mathbf{I}_{2 \times 2} & -\hG_{\bbeta} \\
        \mathbf{0}_{p \times 2} & \hB_{\bbeta}
    \end{pmatrix},
\]
and thus
\begin{align*}
      H^\star B_{\btheta}^{\star -1} = & 
    \begin{pmatrix}
        J(2)-J(1) \\
        \hB_{\bbeta}^{-\top} \EA \left\{ m^\prime(\hbbeta^\top X_{i(2)}) X_{i(2)} - m^\prime(\hbbeta^\top X_{i(1)}) X_{i(1)} \right\} 
    \end{pmatrix}^\top,  \\
    H^\ast B_{\btheta}^{\star -1} U^\ast(D_i) = & \ \hpsi_2(D_i)-\hpsi_1(D_i) + \hmu_2 - \hmu_1 - \delta_0.
\end{align*}
Therefore, 
\begin{equation}
    H^\ast B_{\btheta}^{\star -1} M_{\btheta}^\ast \left( H^\ast B_{\btheta}^{\star -1} \right)^\top = n^{-1} \ES \left\{ \hpsi_2(D_i) - \hpsi_1(D_i) + \hmu_2 - \hmu_1 - \delta_0 \right \}^2 \label{eq:gs-stats-1}.
\end{equation}
Note that $\ES U^\ast(D_i)=\lambda H^{\ast\top}$. Then,
\begin{equation}
    H^\ast B_{\btheta}^{\star -1} \left\{ \ES U^\ast(D_i) \right\} = \lambda H^\ast B_{\btheta}^{\star -1} H^{\ast\top} = n(\hmu_2 - \hmu_1 - \delta_0) \label{eq:gs-stats-2}.
\end{equation}
Replacing (\ref{eq:gs-stats-1}) and (\ref{eq:gs-stats-2}) in (\ref{eq:gs-stats}), we have 
\[
    \hQ_D = \frac{ n ( \hmu_2 - \hmu_1 - \delta_0 )^2 }{ n^{-1} \ES \left\{ \hpsi_2(D_i)-\hpsi_1(D_i) + \hmu_2 - \hmu_1 - \delta_0 \right\}^2 }.
\]
Next, expand the denominator of $\hQ_D$, 
\begin{multline*}
        n^{-1}\ES \left\{\hpsi_2(D_i) - \hpsi_1(D_i) \right\}^2 + 2n^{-1}(\hmu_2 - \hmu_1 - \delta_0) \ES\left\{\hpsi_2(D_i) - \hpsi_1(D_i) \right\}\\ + (\hmu_2 - \hmu_1 - \delta_0 )^2 =  
        n^{-1}\ES \left\{\hpsi_2(D_i) - \hpsi_1(D_i) \right\}^2 + ( \hmu_2 - \hmu_1 - \delta_0 )^2.
\end{multline*}
The second term in the LHS equals zero as $(\hbmu, \hbbeta^\top)$ solves \eqref{eq:ee-mu} and \eqref{eq:ee-beta}, and thus for any $a$, 
\begin{multline*}
        \ES\hpsi_a(D_i) = \left( \sum_{j=1}^n g^\prime(\hbbeta^\top X_{j(a)}) X_{j(a)} \right)^\top \hat{B}_{\bbeta}^{-1} \left( \ES X_i \left\{ Y_i - g(\hbbeta^\top X_i) \right\} \right) \\ + \ES g(\hbbeta^\top X_{i(a)}) - n\hmu_a = 0.
\end{multline*}
So, we can simplify $\hQ_D$ as 
\[
    \frac{ n ( \hmu_2 - \hmu_1 - \delta_0 )^2 }{ \ES \left\{ \hpsi_2(D_i)-\hpsi_1(D_i) \right\}^2 / n + ( \hmu_2 - \hmu_1 - \delta_0 )^2 }.
\]
Finally, applying the delta method to $\hSigma_{\bmu}$ to obtain the variance for $\hmu_2 - \hmu_1$, we have 
\[
    \ES \left\{ \hpsi_2(D_i)-\hpsi_1(D_i) \right\}^2 / n^2 = \hSigma_{\bmu}(2,2) - 2\hSigma_{\bmu}(2,1) + \hSigma_{\bmu}(1,1) = \hsigma_D^2.
\]

~\\
\noindent\textbf{Ratio of means.} Similar to difference in means, define $h(\btheta) = \mu_2-\delta_0\mu_1$. Then $ H^\ast = (J(2)^\top-\delta_0 J(1)^\top, \bzero_p^\top)$ and $\mu_2^\ast=\delta_0\mu_1^\ast$. Thus, $\lambda = n(\hmu_2 - \delta_0\hmu_1)/(\delta_0^2+1)$,
and 
\[
      H^\ast B_{\btheta}^{\star -1} =
    \begin{pmatrix}
        J(2)^\top-\delta_0 J(1)^\top, \EA\left\{ m^\prime( \hbbeta^\top X_{i(2)} ) X_{i(2)} -\delta_0 m^\prime( \hbbeta^\top X_{i(1)}) X_{i(1)} \right\}^\top \hB_{\bbeta}^{-1}
    \end{pmatrix}.
\]
Therefore, 
\begin{align*}
    H^\ast B_{\btheta}^{\star -1}U^\ast(D_i) = & \ \hpsi_2(D_i)-\delta_0\hpsi_1(D_i) + \hmu_2 - \delta_0\hmu_1\\
    \lambda H^\ast B_{\btheta}^{\star -1} H^{\ast\top}
    = & \ n(\hmu_2 - \delta_0\hmu_1).
\end{align*}
After replacing relevant terms in (\ref{eq:gs-stats}), 
\[
    \hQ_R = \frac{ n (\hmu_2 - \delta_0\hmu_1)^2 }{ n^{-1} \ES \left\{ \hpsi_2(D_i) -\delta_0\hpsi_1(D_i) + \hmu_2 - \delta_0\hmu_1\right\}^2 }.
\]
Similar to difference in means, we expand the denominator of $\hQ_R$
\begin{multline*}
    n^{-1}\ES \left\{ \hpsi_2(D_i) - \delta_0\hpsi_1(D_i) \right\}^2 + 2n^{-1}(\hmu_2 - \delta_0\hmu_1) \ES \left\{ \hpsi_2(D_i) - \delta_0\hpsi_1(D_i) \right\}\\ + (\hmu_2 - \delta_0\hmu_1)^2 =  n^{-1}\ES \left\{ \hpsi_2(D_i) - \delta_0\hpsi_1(D_i) \right\}^2  + (\hmu_2 - \delta_0\hmu_1)^2.
\end{multline*}
and then obtain
\[
    \hQ_R = \frac{ n ( \hmu_2 - \delta_0\hmu_1 )^2 }{ \ES \left\{ \hpsi_2(D_i) - \delta_0\hpsi_1(D_i) \right\}^2 / n + ( \hmu_2 - \delta_0\hmu_1 )^2 }.
\]
Finally, applying the delta method to $\hSigma_{\bmu}$ to obtain the variance for $\hmu_2 - \delta_0 \hmu_1$, we have 
\[
    \ES \left\{ \hpsi_2(D_i) - \delta_0\hpsi_1(D_i) \right\}^2 / n^2 = \hSigma_{\bmu}(2,2) - 2\delta_0\hSigma_{\bmu}(2,1) + \delta_0^2\hSigma_{\bmu}(1,1).
\]

\subsection{Generalized score statistics under stratified/biased-coin randomization} \label{si:gscore-stats-strat}

\textbf{Difference in means.} The generalized score statistic can be written as the following sandwich form:
\[
    n^{-1}\times(\ref{eq:gs-stats-2})^\top\times(\ref{eq:gs-stats-3})^{-1}\times(\ref{eq:gs-stats-2}),
\]
where the term in the middle is 
\begin{equation}
     H^\ast B_{\btheta}^{\star -1} M_{\btheta}^{\ast\ast} \left(H^\ast B_{\btheta}^{\star -1}\right)^\top,  \label{eq:gs-stats-3}
\end{equation}
with $M_{\btheta}^{\ast\ast}$ being the variance of $U^\ast(D_i)$ under stratified/biased-coin randomization ($M_{\btheta}^\ast$ is the one under simple randomization). \eqref{eq:gs-stats-3} is exactly the term that \citet{wang2023model} have studied to obtain Theorem 1 in their paper. The only difference is that we assume that $\mu_2-\mu_1 = \delta_0$. Thus, we can directly follow their proofs to calculate $M_{\btheta}^{\ast\ast}$. 

To simplify the notation, we now use $\psi_a$ for $\psi_a^\score$.  Note that the influence function for $\hmu_2 - \hmu_1$ under $H_D$ is $\psi_d^\ast(D_i) \coloneqq \psi_2(D_i)-\psi_1(D_i) + \mu_2 - \mu_1 - \delta_0$. Write $\psi_d^\ast(D_i) = \sum_{a=1}^k I(A_i=a) \psi_d^{a\ast}(D_i)$, where $\psi_d^{a\ast}(D_i) \coloneqq \psi_d^\ast(Y_i(a),a,W_i)$. Following the proofs in Theorem 1 in the supplementary material of \citet{wang2023model}, 
\begin{align*}
    & \eqref{eq:gs-stats-3} \to \sum_{a=1}^2 \pi_a \E \left[ \Var \left\{ \left. \psi_d^{a\ast}(D_i) \right\vert S_i \right\} \right] + \E \left[ \left( \E \left\{ \left. \sum_{a=1}^2 \pi_a \psi_d^{a\ast}(D_i) \right\vert S_i \right\} \right)^2 \right]\\
    & \eqref{eq:gs-stats-1} \to \sum_{a=1}^2 \pi_a \E\left\{ \left( \psi_d^{a\ast}(D_i) \right)^2 \right\} = \sum_{a=1}^2 \pi_a \left( \E \left[ \Var \left\{ \left. \psi_d^{a\ast}(D_i) \right\vert S_i \right\} \right]  + \E \left[ \left( \E \left\{ \left. \psi_d^{a\ast}(D_i) \right\vert S_i \right\} \right)^2 \right] \right) \\
    & \eqref{eq:gs-stats-1} - \eqref{eq:gs-stats-3} = \ \pi_1\pi_2 \E \left[ \left( \E \left\{ \left. \psi_d^{2\ast} (D_i) - \psi_d^{1\ast}(D_i) \right\vert S_i \right\} \right)^2 \right] \\
    & = \ \pi_1\pi_2 \E \left[ \left( \E \left \{ \left. (\frac{ I(A_i=2) }{\pi_2}-\frac{I(A_i=1)}{\pi_1})(\psi_2(D_i)-\psi_1(D_i) + \mu_2 - \mu_1 - \delta_0) \right\vert S_i \right\} \right)^2 \right]. 
\end{align*}
It is straightforward to verify than \eqref{eq:gs-stats-1} - \eqref{eq:gs-stats-3} is exactly the same (with slightly different notations) as the additional term for stratified/biased-corn randomization in Theorem 1 of \cite{wang2023model}. 

Combining the above, the generalized score statistic under stratified/biased-coin randomization is 
\[
    \frac{n(\hmu_2 - \hmu_1 - \delta_0)^2}{n^{-1}\ES\big\{\hpsi_2(D_i)-\hpsi_1(D_i) + \hmu_2 - \hmu_1 - \delta_0\big\}^2 - [\text{empirical estimate of }(\ref{eq:gs-stats-1})-(\ref{eq:gs-stats-3})]}.
\]
As \eqref{eq:gs-stats-1} - \eqref{eq:gs-stats-3} is always non-negative, the above statistic is always no smaller than $\hQ_D$. 

~\\
\noindent\textbf{Ratio of means.} Note that the influence function for $\hmu_2 / \hmu_1$ under $H_R$ is $\psi_r^\ast(D_i) \coloneqq \psi_2(D_i)-\delta_0\psi_1(D_i) + \mu_2 - \delta_0\mu_1$. Similar to difference in means, we can obtain its generalized score statistic, which is equal to $\hQ_R$ minus an additional term in its denominator. This additional term, which can be derived by replacing $\psi_d^{a\ast}(D_i)$ with $\psi_r^{a\ast}(D_i)$ in \eqref{eq:gs-stats-1} and \eqref{eq:gs-stats-3}, is written as
\[
    \pi_1\pi_2 \E \left[ \left( \E \left \{ \left. (\frac{ I(A_i=2) }{\pi_2}-\frac{I(A_i=1)}{\pi_1})(\psi_2(D_i)-\delta_0\psi_1(D_i) + \mu_2 - \delta_0\mu_1) \right\vert S_i \right\} \right)^2 \right]. 
\]

\clearpage 

\section{R Code Demo} \label{si:code}

\begin{lstlisting}

# R code demo for g-computation, 
# incuding three variance estimators, Wald/score tests 
# and corresponding confidnce intervals

library(dplyr)

var_gcomp_glm_mest <- function(mu, g, wm, muDbeta) {  
  
  # Variance estimator based on M-estimation 
  # mu: gcomp per arm (k arms)
  # g: predicted values of each subject for each arm (dim=k,n)
  # wm: fitted working model
  # muDbeta: derivative of mu per arm w.r.t. beta (dim=k,p)
  
  # influence function for beta
  betaPhi <- (vcov(wm) * nobs(wm)) %*% t(wm$x * (wm$y - wm$fitted.values))  
  # influence function for mu
  muPhi <- muDbeta %*% betaPhi + g - mu  # dim=(k,n)
  # variance for mu
  Sigma <- var(t(muPhi)) / nobs(wm)
  
  return(list(var = Sigma, phi = muPhi))
  
}


var_gcomp_glm_sp1 <- function(mu, g, wm, piA) {  
  
  # Variance estimator based on efficient influence functions
  # mu: gcomp per arm (k arms)
  # g: predicted values of each subject for each arm (dim=k,n)
  # wm: fitted working model
  # piA: true/emprical propotions for k arms
  
  muPhi <- t(sapply(1:nrow(g), function(a)  # dim=(k,n)
    ifelse(wm$x[, a] == 1, 1, 0) / piA[a] * (wm$y - g[a, ]) + g[a, ] - mu[a]))  
  # variance for mu
  Sigma <- var(t(muPhi)) / nobs(wm)
  
  return(list(var = Sigma, phi = muPhi))
  
}


var_gcomp_glm_sp2 <- function(mu, g, wm, piA) {
  
  # Alternative Variance estimator based on 
  #   efficient influence functions proposed by Ting Ye
  # mu: gcomp per arm (k arms)
  # g: predicted values of each subject for each arm (dim=k,n)
  # wm: fitted working model
  # piA: true/emprical propotions for k arms
  
  # variance for mu

  # v1 <- diag(sapply(1:k, function(a)    # based on original formula
  #         var(wm$y[wm$x[, a] == 1] - g[a, wm$x[, a] == 1) / piA[a])) 
  
  # decomposed (v2 and v3) to align with the implementation of RobinCar package 
  v2 <- sapply(1:nrow(g), function(a) 
    cov(wm$y[wm$x[, a] == 1], t(g[, wm$x[, a] == 1])))
  v3 <- var(t(g))
  v1 <- diag(sapply(1:nrow(g), function(a) 
    var(wm$y[wm$x[, a] == 1]) / piA[a])) +  
    (diag(diag(v3)) - 2 * diag(diag(v2))) * diag(1/piA)
  
  Sigma <- (v1 + v2 + t(v2) - v3) / nobs(wm)
  
  return(list(var = Sigma))
  
}


score_test_diff <- function(mu, Sigma, wm, null, level) {
  
  # mu: gcomp for control and tested arms
  # Sigma: variance of control and tested arms gcomp estimators
  # wm: working model
  # null: null value
  # level: confidence level
  
  v <- Sigma[1, 1] - 2 * Sigma[1, 2] + Sigma[2, 2]
  z <- (mu[2] - mu[1] - null) / sqrt(v + (mu[2] - mu[1] - null)^2 / nobs(wm))
  pval <- 1 - pnorm(z)  # 1-sided
  ci <- mu[2] - mu[1] + sqrt(v * qchisq(level, df = 1) / 
                                 (1 - qchisq(level, df = 1) / nobs(wm))) * c(-1, 1)

  return(list(statistic = z, p.value = pval, conf.int = ci))
  
}


score_test_ratio <- function(mu, Sigma, wm, null, level) {
  
  # mu: gcomp for control and tested arms
  # Sigma: variance of control and tested arms gcomp estimators
  # wm: working model
  # null: null value
  # level: confidence level
  
  z <- (mu[2] - mu[1] * null) / 
    sqrt(Sigma[1, 1] * null^2 - 2 * Sigma[1, 2] * null + Sigma[2, 2] + 
           (mu[2] - mu[1] * null)^2 / nobs(wm))
  pval <- 1 - pnorm(z)  # 1-sided
  
  a <- (1 - qchisq(level, df = 1) * (Sigma[1, 2] / (mu[1] * mu[2]) + 1 / nobs(wm))) / 
    (1 - qchisq(level, df = 1) * (Sigma[1, 1] / mu[1]^2 + 1 / nobs(wm)))         # for ci
  b <- (1 - qchisq(level, df = 1) * (Sigma[2, 2] / mu[2]^2 + 1 / nobs(wm))) / 
    (1 - qchisq(level, df = 1) * (Sigma[1, 1] / mu[1]^2 + 1 / nobs(wm)))         # for ci
  ci <- mu[2] / mu[1] * (a + sqrt(a^2 - b) * c(-1, 1)) 
  
  return(list(statistic = z, p.value = pval, conf.int = ci))
  
}


# simulation for a hypothetical trial of N =326 with pi1 = 30% and pi2 = 45%
 
betaZ <- rep(sqrt(log(2)^2/3), 3); exp(sqrt(sum(betaZ^2)))
betaA <- c(-0.9355, -0.2224)

set.seed(12345)

n <- 326

df_sim <- data.frame(
  A = rep(1, n),
  W1 = rnorm(n, 0, 1),
  W2 = rnorm(n, 0, 1),
  W3 = rnorm(n, 0, 1),
  Y = rep(0, n)
) 

df_sim$A[sample(n, size = n/2, replace = FALSE)] <- 2
df_sim$Y <- mapply(function(a, z1, z2, z3) {
  rbinom(n = 1, size = 1, prob = plogis(betaA[a]+crossprod(betaZ,c(z1,z2,z3))))
}, df_sim$A, df_sim$W1, df_sim$W2, df_sim$W3)

## unadjusted

p_unadj <- tapply(df_sim$Y, df_sim$A, mean)   # marginal prop of Y = 1

nA <- table(df_sim$A)
var_p_unadj <- p_unadj * (1 - p_unadj) / nA

unadj <- p_unadj[2] - p_unadj[1]  # risk difference
var_unadj <- sqrt(sum(var_p_unadj))


## g-computation
# estimation 

wm <- glm(Y ~ 0 + factor(A) + W1 + W2 + W3, family = binomial, data = df_sim, x = TRUE)
# wm <- glm(Y ~ 0 + factor(A) + W1 + W2 + W3, family = poisson, data = df_sim, x = TRUE)

g_predict <- sapply(1:2, function(a) 
  predict(wm, newdata = df_sim %>% mutate(A = a), type = "response"))
p_gc <- colMeans(g_predict)
diff_gc <- p_gc[2] - p_gc[1]

# variance 

if (wm$family$link == "logit") {      # for M-estimation 
  
  g_deriv <- sapply(1:2, function(a) { x <- wm$x; x[, 1:2] <- 0; x[, a] <- 1; 
  colMeans(g_predict[, a] * (1 - g_predict[, a]) * x)})
  
} else {
  
  g_deriv <- sapply(1:2, function(a) { x <- wm$x; x[, 1:2] <- 0; x[, a] <- 1; 
  colMeans(g_predict[, a] * x)})
  
}

mu_gc_var <- var_gcomp_glm_mest(p_gc, t(g_predict), wm, t(g_deriv))
# mu_gc_var <- var_gcomp_glm_sp1(p_gc, t(g_predict), wm, nA/n)
# mu_gc_var <- var_gcomp_glm_sp2(p_gc, t(g_predict), wm, c(0.5, 0.5))

# inference 

score_test_diff(p_gc, mu_gc_var$var, wm, 0, 0.95) 


# p_gc[2] / p_gc[1]   # ratio
# score_test_ratio(p_gc, mu_gc_var$var, wm, 1, 0.95)

\end{lstlisting}

\end{document}